\documentclass[lettersize,journal]{IEEEtran}
\usepackage{amsmath,amsfonts}
\usepackage{algorithmic}
\usepackage[ruled,noline]{algorithm2e}
\usepackage{array}
\usepackage{tabularx}
\usepackage{makecell}  
\usepackage{textcomp}
\usepackage[caption=true,font=footnotesize,textfont=rm]{subfig}
\usepackage{subfloat}
\usepackage{textcomp}
\usepackage{tikz}

\usepackage{amssymb}
\usepackage{bbding}
\usepackage{pifont}
\usepackage{url}
\usepackage{fancyhdr}
\usepackage{hyperref}
\usepackage{verbatim}
\usepackage{graphicx}
\usepackage{multirow}
\usepackage{cite}
\usepackage{booktabs}
\usepackage{ulem}

\usepackage{float}
\usepackage{makecell}
\usepackage{orcidlink}
\usepackage{tikz}
\usepackage[switch]{lineno}
\begin{document}
\title{Speech-driven Personalized Gesture Synthetics: Harnessing Automatic Fuzzy Feature Inference}

\author{Fan Zhang\hspace{-1.5mm}$^{~\orcidlink{0000-0002-9534-1777}}$, 
Zhaohan Wang\hspace{-1.5mm}$^{~\orcidlink{0009-0005-4783-6213}}$, 
Xin Lyu \hspace{-1.5mm}$^{~\orcidlink{0009-0000-1055-6334}}$,
Siyuan Zhao\hspace{-1.5mm}$^{~\orcidlink{0009-0009-5831-8699}}$,
Mengjian Li \hspace{-1.5mm}$^{~\orcidlink{0009-0000-9698-8508}}$,
Weidong Geng \hspace{-1.5mm}$^{~\orcidlink{0000-0002-2709-396X}}$,
Naye Ji 
% $^{(\textrm{\Letter})}$
\hspace{-1.5mm}$^{~\orcidlink{0000-0002-6986-3766}}$,
Hui Du\hspace{-1.5mm}$^{~\orcidlink{0000-0001-6737-9420}}$, 
Fuxing Gao\hspace{-1.5mm}$^{~\orcidlink{0009-0008-5586-4734}}$, 
Hao Wu \hspace{-1.5mm}$^{~\orcidlink{0009-0006-2083-3001}}$, 
Shunman Li \hspace{-1.5mm}$^{~\orcidlink{0009-0008-3269-0145}}$%
% ~\IEEEmembership{Staff,~IEEE,}
\thanks{Fan Zhang is with the Faculty of Humanities and Arts, Macau University of Science and Technology, Macau, China; The College of Media Engineering, Communication University of Zhejiang, China; Research Center for Artificial Intelligence and Fine Arts, Zhejiang Lab, Zhejiang, China (e-mail: fanzhang@cuz.edu.cn)}
\thanks{Zhaohan Wang, Xin Lyu are with the School of Animation and Digital Arts
Communication University of China, Beijing, China (e-mail: 2022 201305j6018@cuc.edu.cn; lvxinlx@cuc.edu.cn) }
\thanks{Mengjian Li is with the Research Center for Artificial Intelligence and Fine Arts, Zhejiang Lab, Zhejiang, China(e-mail: limengjian@zhejianglab.com)}
\thanks{Weidong Geng is with the College of Computer Science and Technology, Zhejiang University, the Research Center for Artificial Intelligence and Fine Arts, Zhejiang Lab, Zhejiang, China (e-mail: gengwd@zju.edu.cn)}
\thanks{Naye Ji, Hui Du, Fuxing Gao, Hao Wu are with the College of Media Engineering, Communication University of Zhejiang, China (e-mail: jinaye@cuz.edu.cn; fuxing@cuz.edu.cn; duhui@cuz.edu.cn; 210207140@stu.cuz.edu.cn)}
\thanks{Siyuan Zhao is with the Faculty of Humanities and Arts, Macau University of Science and Technology, Macau, China(e-mail: 2109853jai30001@stu dent.must.edu.mo)}

\thanks{Shunman Li is with Zhejiang Institute of Economics and Trade, Zhejiang, China (e-mail: 2017000018@zjiet.edu.cn)}}

        % <-this % stops a space



\maketitle

\begin{abstract}
Speech-driven gesture generation is an emerging field within virtual human creation. However, a significant challenge lies in accurately determining and processing the multitude of input features (such as acoustic, semantic, emotional, personality, and even subtle unknown features). Traditional approaches, reliant on various explicit feature inputs and complex multimodal processing, constrain the expressiveness of resulting gestures and limit their applicability. To address these challenges, we present \textit{Persona-Gestor}, a novel end-to-end generative model designed to generate highly personalized 3D full-body gestures solely relying on raw speech audio. The model combines a fuzzy feature extractor and a non-autoregressive Adaptive Layer Normalization (AdaLN) transformer diffusion architecture. The fuzzy feature extractor harnesses a fuzzy inference strategy that automatically infers implicit, continuous fuzzy features. These fuzzy features, represented as a unified latent feature, are fed into the AdaLN transformer. The AdaLN transformer introduces a conditional mechanism that applies a uniform function across all tokens, thereby effectively modeling the correlation between the fuzzy features and the gesture sequence. This module ensures a high level of gesture-speech synchronization while preserving naturalness. Finally, we employ the diffusion model to train and infer various gestures. Extensive subjective and objective evaluations on the Trinity, ZEGGS, and BEAT datasets confirm our model's superior performance to the current state-of-the-art approaches. \textit{Persona-Gestor} improves the system's usability and generalization capabilities, setting a new benchmark in speech-driven gesture synthesis and broadening the horizon for virtual human technology. Supplementary videos and code can be accessed at \href{https://zf223669.github.io/Diffmotion-v2-website/} {https://zf223669.github.io/Diffmotion-v2-website/.}

\end{abstract}


\begin{figure}[!t]
\includegraphics[width=0.49\textwidth]{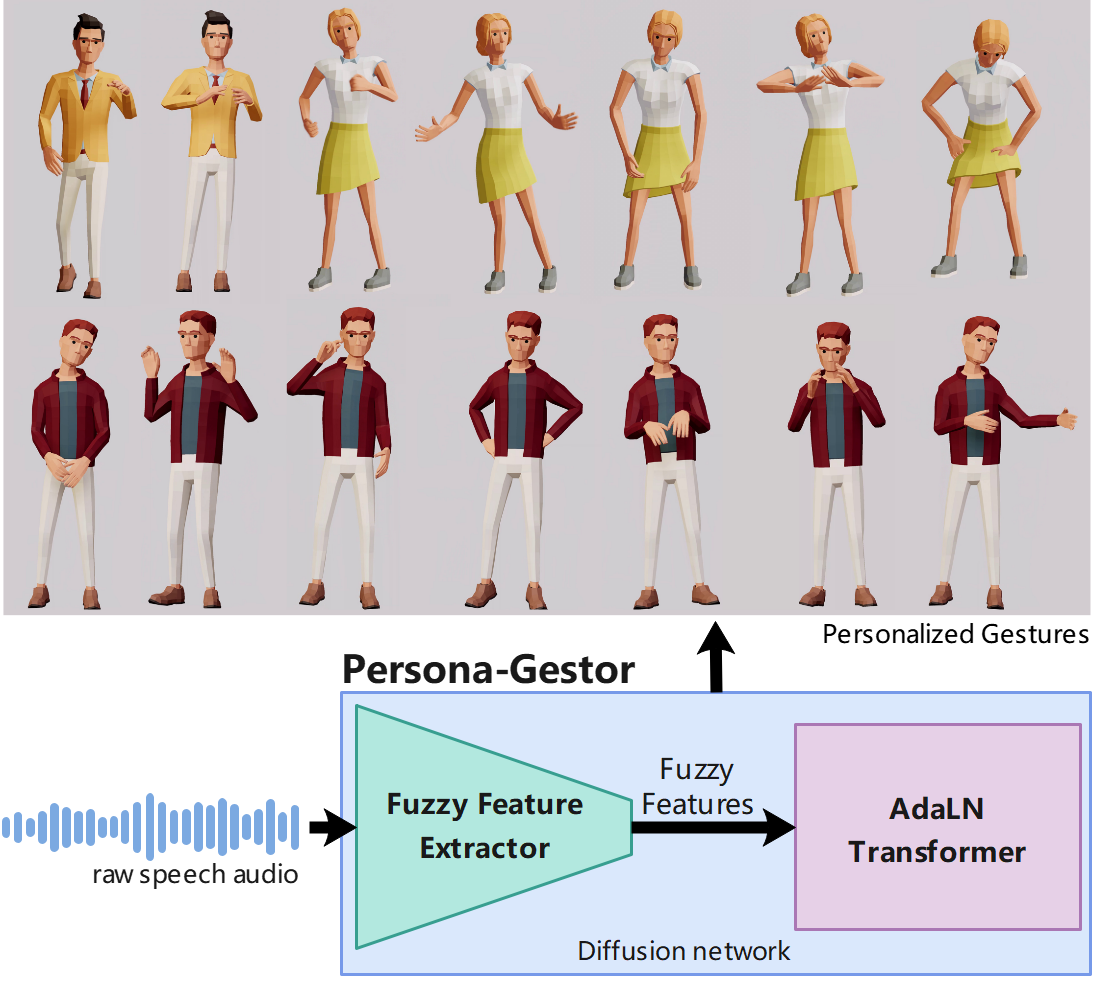}
\begin{center}
\caption{Each pose depicted is personalized gestures generated solely relying on raw speech audio. Persona-Gestor offers a versatile solution, bypassing complex multimodal processing and thereby enhancing user-friendliness. }
\label{fig:result_showcase} 
\end{center}
\end{figure}

\section{Introduction}
\IEEEPARstart{R}{ecent} advancements have significantly expanded the use of 3D virtual human technology. Its growing appeal spans numerous applications, including animation, gaming, digital receptionists, and human-computer interaction. A major task in this research area is to create credible, personalized co-speech gestures. Speech-driven gesture generation through deep learning provides a cost-effective solution, eliminating the need for manual intervention associated with conventional motion capture systems. 

However, the primary challenges in speech-driven gesture generation face precisely identifying the vast array of input conditions necessary for driving gesture synthesis. This complexity arises because co-speech gestures are shaped by an extensive range of factors, including acoustics, semantics, emotions, personality traits, and demographic variables like gender, age, etc.

Previous approaches\cite{bhattacharya2021speech2affectivegestures,yang_diffusestylegesture_2023,yang2023DiffuseStyleGestureaplus,alexanderson2023listen, yang_unifiedgesture_2023, li2023Audio2Gesturesa,ghorbani2023ZeroEGGS} have explored the use of manual labels and diverse feature inputs to facilitate the synthesis of personalized gestures. Nonetheless, these methods depend heavily on various unstructured feature inputs and require complex multimodal processing. These approaches present a significant barrier to the practical application and broader adoption of virtual human technologies.

The fuzzy inference strategy, which pertains to the concept of fuzzy logic\cite{zadeh1965fuzzy}, is particularly useful in the field for dealing with uncertain or imprecise information. The fuzzy inference strategy is known for its effectiveness in speech-emotion recognition\cite{vashishtha2020unsupervised} and audio classification\cite{patil2019content}. These methods do not necessarily require explicit classification outputs but instead provide fuzzy feature information, which broadens the explicit discrete space into an expansive implicit continuous fuzzy space. The information in fuzzy space better aligns with the actual scenario. Relevantly, research in psychology highlights the significance of various factors in speech\cite{calvo2015oxford,goudbeek2010beyond,hirschberg2015advances,campbell2011intersecting}. These factors, called fuzzy features, are intricately intertwined with co-speech gestures. These studies present novel opportunities for synthesizing personalized gestures based solely on speech audio, thereby simplifying the feature inputs and reducing the complexity of multimodal processing.

Another challenge in this field is ensuring a high level of gesture-speech synchronization while preserving naturalness. Recent developments have focused on the application of Transformer and Diffusion-based models. This methodological shift has led to substantial progress in the efficiency and flexibility of gesture-generation technologies. Key examples of such innovative efforts include Taming\cite{zhu_taming_2023}, Diffuse Style Gesture\cite{yang_diffusestylegesture_2023}, Diffuse Style Gesture+\cite{yang2023DiffuseStyleGestureaplus}, GestureDiffuClip\cite{ao2023GestureDiffuCLIP}, and LDA\cite{alexanderson2023listen}. Yet, these approaches encounter challenges with either insufficient or excessive correlation between gesture and speech, reducing the naturalness of the generated gestures.

The success of the Diffusion Transformers(DiTs) in text2image\cite{peebles2022Scalable} and text2video generation tasks, such as Sora \footnote{https://openai.com/sora}, which incorporates AdaLN, marks a significant advancement. This framework introduces a conditional mechanism that applies a uniform function across all tokens, enhancing the model's ability to represent conditional and output features. This conditional mechanism also holds promise for effectively enhancing the ability to model the intricate mapping between speech and gestures. While the original DiTs take discrete text prompts as conditional inputs, its adaptability for sequence-to-sequence tasks, such as speech-driven gesture generation, presents an area of exploration. 

In this study, we propose \textit{Persona-Gestor}, a novel approach aimed at synthesizing personalized gestures solely from raw speech audio. This model innovatively introduces a fuzzy feature inference strategy within its condition extractor and incorporates AdaLN in a diffusion-based transformer module. \textit{Persona-Gestor} transitions from explicit conditions to a nuanced, continuous representation of fuzzy features by employing fuzzy inference, which captures a broad spectrum of stylistic nuances and specific audio details. These features are integrated into a unified latent representation, synthesizing intricate 3D full-body gestures. Adopting AdaLN significantly enhances the model's capability to depict the nuanced relationship between speech and gestures. Leveraging a diffusion process, the framework can generate diverse gesture outputs, showcasing the potential for high fidelity in gesture synthesis.

For clarity, our contributions are summarized as follows:
\begin{itemize}
% This strategy implicitly extracts personalized fuzzy features autonomously, paving the way for an end-to-end framework that simplifies handling unstructured multimodal features. This approach improves usability and system generalization and represents a pioneering step in utilizing fuzzy inference for the exclusive synthesis of personality gestures.
% implicitly extracting and unifying fuzzy features as a condition
\item{\textbf{We pioneering introduce the fuzzy feature inference strategy that enables driving a wider range of personalized gesture synthesis from speech audio alone, removing the need for style labels or extra inputs.} This fuzzy feature extractor improves the usability and the generalization capabilities of the system. To the best of our knowledge, it is the first approach that uses fuzzy features to generate co-speech personalized gestures.}

\item{\textbf{We combined AdaLN transformer architecture within the diffusion model to enhance the Modeling of the gesture-speech interplay.} We demonstrate that this architecture can generate gestures that achieve an optimal balance of natural and speech synchronization.}

\item{\textbf{Extensive subjective and objective evaluations reveal our model superior outperform to the current state-of-the-art approaches.} These results show the remarkable capability of our method in generating credible, speech-appropriateness, and personalized gestures.}
\end{itemize}

\section{RELATED WORK}\label{sec:RELATED WORK}
The present discussion offers a succinct overview of the conditional extraction mechanism and generative models within speech-driven gesture generation.

\subsection{Condition Extraction Mechanism}
Recent advancements in co-speech gesture generation systems have incorporated various unstructured conditional information as input.  

Selecting optimal representations for conditional input is a crucial research challenge in creating virtual human motions\cite{windle2022UEA}\cite{yu2021multimodal}. For accurate reflection of gestures that match the auditory perception, prevalent research\cite{alexanderson_simon_style-controllable_2020,taylorsarah2021SpeechDriven,zhang2023diffmotion,ghorbani2023ZeroEGGS} utilizes preprocessed audio features, such as MFCCs, log amplitude spectrogram, etc. Li et al.\cite{li2023Audio2Gesturesa} develop a model for direct audio-to-gesture mapping. Despite these methods capturing acoustic nuances, the quest for richer feature sets continues. This has prompted investigations into the WavLM model, a refined, pre-trained wav2vec framework, for enhanced speech extraction, showcasing in ReprGesture\cite{yang2022ReprGesturea}, QPGesture\cite{yang2023QPGesture}, and DiffuseStyleGesture\cite{yang_diffusestylegesture_2023}.

Text-based co-speech gesture synthesis has seen significant contributions, such as Yoon et al.'s\cite{yoonyoungwoo2019Robots} recurrent neural network approach and Taras et al.'s\cite{kucherenkotaras2020Gesticulator} system, which merges acoustic and semantic speech features, employing BERT for semantic analysis \cite{devlinjacob2018Bert}. Additionally, Uttaran et al.\cite{bhattacharyauttaran2021Text2Gestures} utilize GloVe embeddings\cite{penningtonjeffrey2014Glove} to surpass models of similar dimensions, like Word2Vec\cite{mikolovtomas2013Distributed} and FastText\cite{bojanowskipiotr2017Enriching}. Merging acoustic with semantic data offers a valuable path to enhance the relevance and context of generated gestures. Nonetheless, these modalities' manual alignment and integration pose a challenge in effectively superior gesture synthesis.

For creating style-specific gestures, ReprGesture\cite{yang2022ReprGesturea} and QPGesture\cite{yang2023QPGesture} integrate textual data with audio features, whereas DiffuseStyleGesture\cite{yang_diffusestylegesture_2023} employs discrete labels to influence the stylistic aspects of the gestures produced. LDA\cite{alexanderson2023listen} enables the system to generate style gestures with classifier-free guidance. Additionally, recent research has explored using textual prompts to generate stylized gestures\cite{ao2023GestureDiffuCLIP}. Given that human emotions are more accurately represented on a continuous spectrum\cite{A_Circumplex_Model_of_Affect}\cite{cambria2012hourglass} and emerge from a complex interplay of fuzzy factors, depending on discrete emotion labels can overly simplify the gesture generation process. This could limit the expressiveness and subtlety of the produced gestures. To address these limitations, Ghrobani et al.\cite{ghorbani2023ZeroEGGS} introduced ZeroEGGS, a model that utilizes example motion clips to guide the style of gestures. Although achieving zero-shot is feasible, it still necessitates sample animation clips.

\subsection{Generative approaches} 
DiffMotion\cite{zhang2023diffmotion}, is the pioneering application of diffusion models integrating an LSTM for the synthesis of diverse gestures. UnifiedGesture\cite{yang_unifiedgesture_2023} presents a retargeting network to learn latent homeomorphic graphs to homeomorphic graphs for various gesture representations. Maximizing the transformer architecture's potential, Alexanderson et al.\cite{alexanderson2023listen} enhanced DiffWave by replacing dilated convolutions. Conformers\cite{gulati2020Conformer} implementing classifier-free guidance to improve style expression.  GestureDiffuCLIP\cite{ao2023GestureDiffuCLIP} propose a network based on the transformer and AdaIN layers to incorporate style guidance into the diffusion model. LivelySpeaker\cite{zhi_livelyspeaker_2023} depends on contrastive learning to create a joint embedding space between gestures and transcripts. DiffuseStyleGesture (DSG)\cite{yang_diffusestylegesture_2023} and DSG+\cite{yang2023DiffuseStyleGestureaplus}, integrating cross-local attention and layer normalization within transformers. Conversely, these methodologies face difficulties in achieving an optimal balance between gesture and speech synchronization, resulting in gestures that may appear either underrepresented or overly matched.

In this study, we employ a fuzzy feature inference strategy to implicitly capture fuzzy features in speech audio, synthesizing natural, personalized co-speech gestures solely relying on raw speech audio without additional modalities. Furthermore, we employ an AdaLN transformer architecture to enhance the model's capacity to capture the intricate relationship between speech and gestures. 
%%%%%%%%%%%%%%%%% Figure 1 %%%%%%%%%%%%%%%%%%%%%%%%%%%%
\begin{figure*}[!h]
\centering
\subfloat[Overall schematic]{\includegraphics[width=0.34\textwidth]{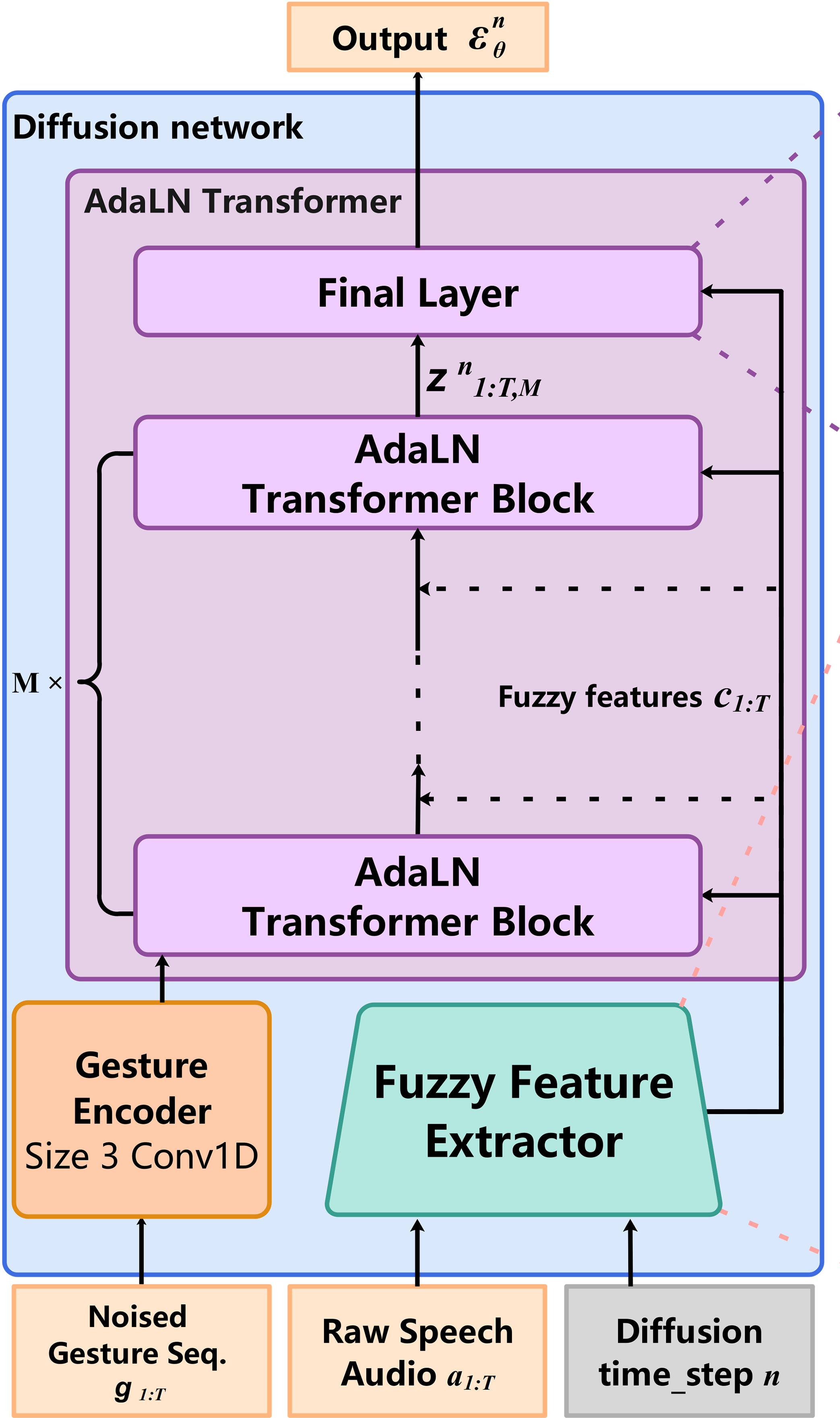}%
\label{fig:architecture1}}
\hfil
\subfloat[Fuzzy Feature Extractor (below), Final Layer in AdaLN Transformer (above)]{\includegraphics[width=0.32\textwidth]{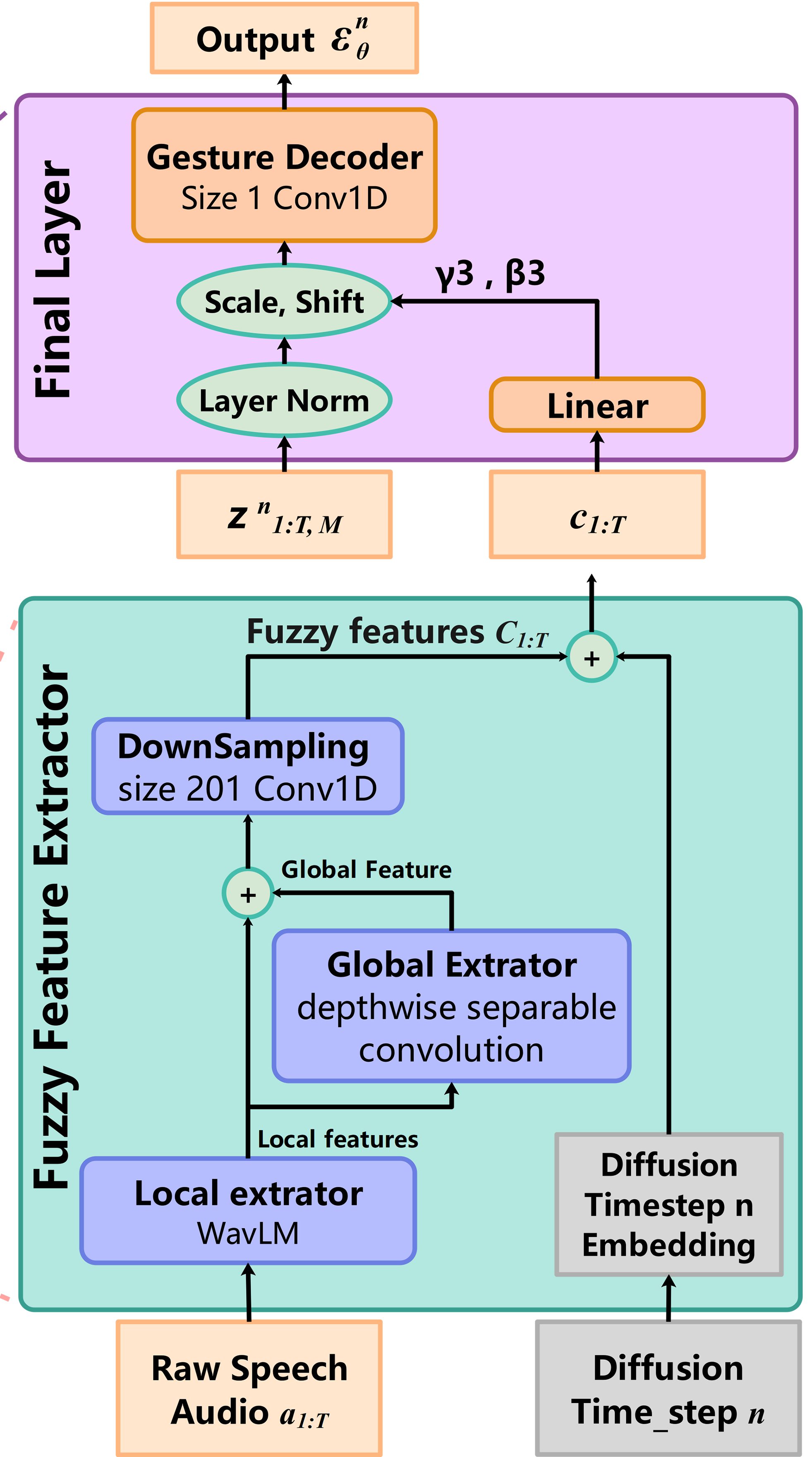}%
\label{fig:architecture2}}
\hfil
\subfloat[AdaLN Transformer Block]{\includegraphics[width=0.311\textwidth]{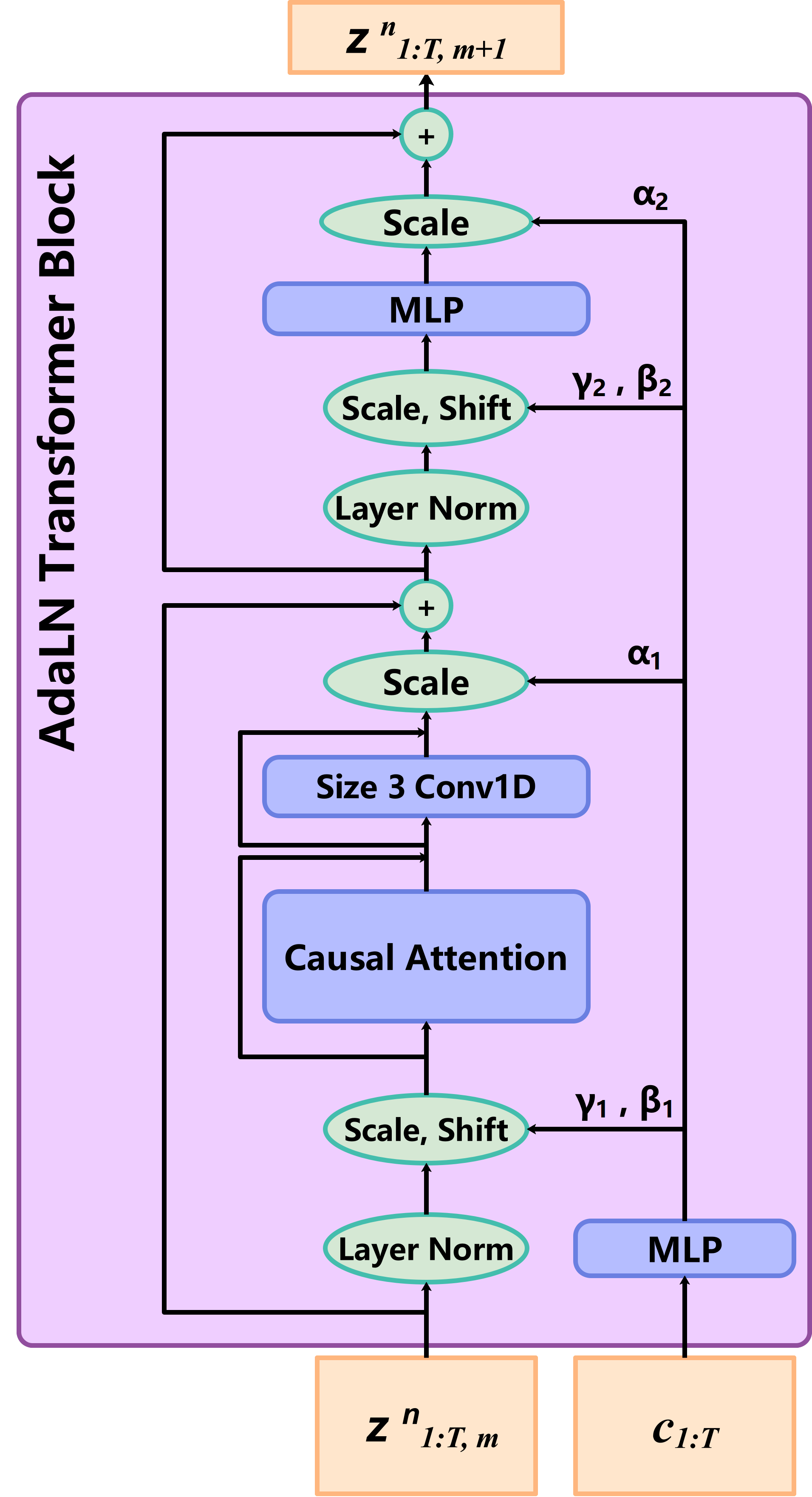}%
\label{fig:architecture3}}
\caption{The Architecture of Persona-Gestor mainly integrates a fuzzy feature extractor and an adaptive layer normalization (AdaLN) transformer diffusion architecture. The fuzzy feature extractor comprises a dual-component framework to comprehensively capture the fuzzy style and detail-oriented audio features. These features, as unified latent features, are subsequently fed into the AdaLN transformer to model the relationship with the accompanist gesture, facilitating the estimation of diffusion noise for the diffusion model. (a) Overall Schematic. (b) Fuzzy Feature Extractor. (c) AdaLN Transformer Block.}
\label{fig:architecture}
\end{figure*}
%%%%%%%%%%%%%%%%% Figure 1 %%%%%%%%%%%%%%%%%%%%%%%%%%%%

\section{System Overview}\label{sec:PROPOSED APPROACH}
Persona-Gestor, as an end-to-end architecture, processes raw speech audio as its sole input, synthesizing personalized gestures that adeptly balance naturalness with synchronized alignment to speech.  
% %%%%%%%%%%%%%%%%% poses, gestures, identity

\subsection{Problem Formulation}
We introduce the challenge of co-speech gesture generation by framing it as a sequence-to-sequence problem, where the objective is to translate a sequence of speech audio features into a corresponding sequence of gestures. We denote the sequence of full-body gesture features and the sequence of the audio signal as \(g^0 = g^0_{1:T} \in [g^0_1,...,g^0_t,...,g^0_T] \in \mathbb{R}^{T \times (D+3+3)}\)  and \(a = a_{1:T} \in [a_1,...,a_t,...,a_T] \in \mathbb{R}^{T}\). \(g^0_t = \mathbb{R}^{(D+3+3)}\) symbolizes the representation of 3D joint angles, along with the root positional and rotational velocity at frame \(t\), where \(D\) denoting the number of channels for these joints. The superscript indicates the diffusion time step \(n\). Here, \(a_t\) refers to the current subsequence audio waveform signal at frame \(t\), while \(T\) denotes the sequence length. Let us define  \(p_\theta(\cdot)\) as the Probability Density Function (PDF), which aims to approximate the actual distribution of gesture data \(p(\cdot)\) and enables easy sampling. The objective is to generate a non-autoregressive whole pose sequence (\(g^0\)) from its conditional probability distribution given audio signal (\(a\)) as covariate:
\begin{equation}
\begin{aligned}
g^0 \sim p_\theta\left(g^0|a\right) &\approx p(\cdot) := p\left(g^0|a\right) 
\end{aligned}\label{eq:qxc}
\end{equation}

where the \(p_\theta(\cdot)\) aims to approximate \(p(\cdot)\) trained by the denoising diffusion model.

\subsection{Model Architecture}
The architecture of Persona-Gestor is depicted in Figure \ref{fig:architecture}. It comprises four primary components: (1) a Fuzzy Feature Extractor, (2) an AdaLN Transformer, (3) a Gesture Encoder and Decoder,  and (4) a diffusion network.

\subsubsection{Fuzzy Feature Extractor}
\begin{figure}[htbp]
\includegraphics[width=0.45\textwidth]{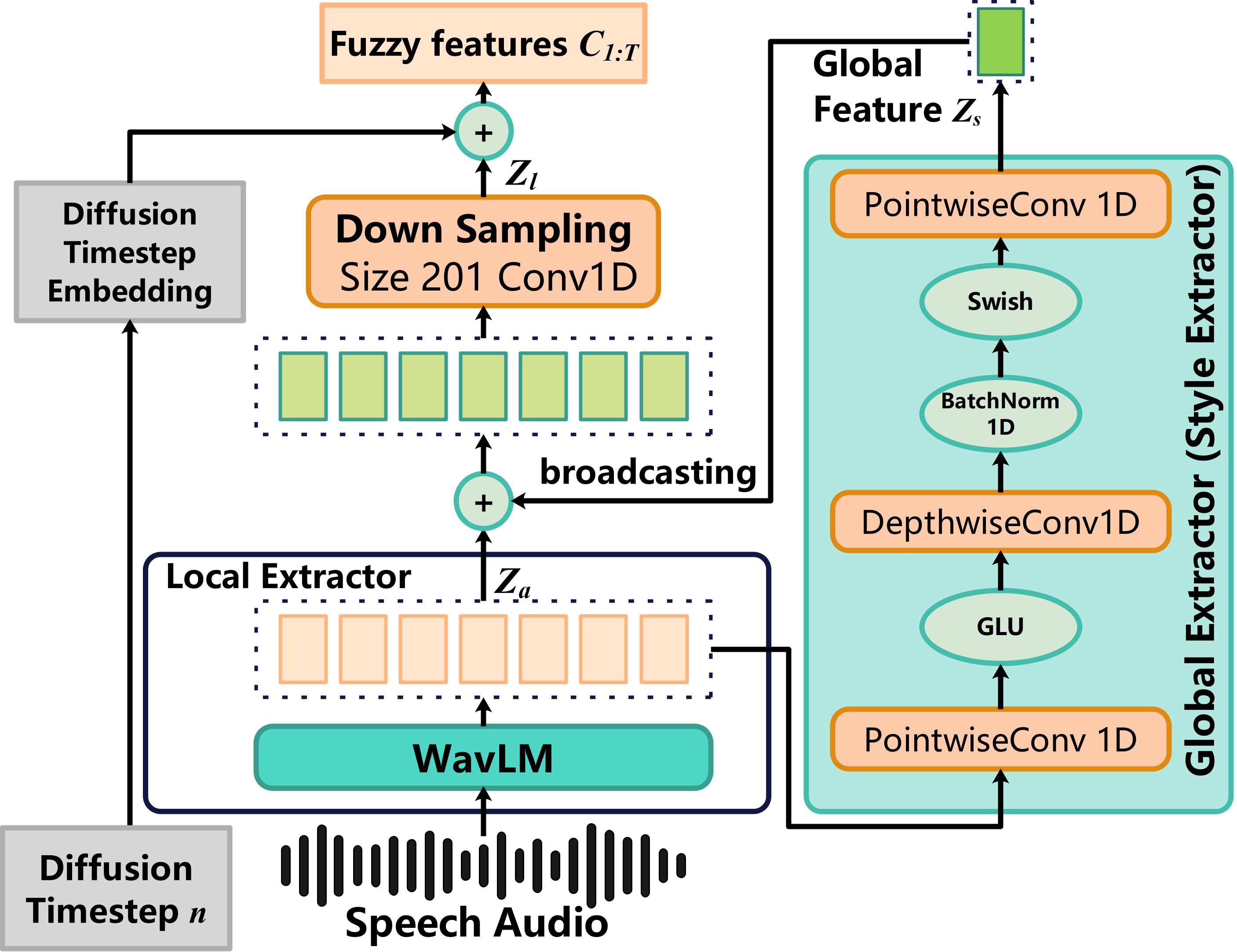}
\begin{center}
\caption{An overview of the fuzzy inference condition extractor. }
\label{fig:ConditionEncoder} 
\end{center}
\end{figure}

This module utilizes a fuzzy inference strategy, meaning it does not generate explicit classification outputs. Instead, it offers implicit, continuous, fuzzy feature information, automatically learning and inferring the global style and details directly from raw speech audio. The module, showcased in Figure \ref{fig:architecture2} and Figure \ref{fig:ConditionEncoder}, is a dual-component extractor that integrates both global and local extractors. The local extractor leverages the WavLM large-scale pre-trained model\cite{chen2022Wavlm} to convert the audio sequence into tokens. We chose WavLM for its adeptness at extracting the complex features of speech audio to capture universal audio latent representations, denoted as $z_a$.

We observe that the local extractor alone falls short of fully capturing the array of stylistic features and ensuring style consistency across sequences. To overcome this, we integrate a global style extractor, employing a depthwise separable convolution 1D layer\cite{kaiser_depthwise_2017} across the $z_a$. This global extractor is designed to automatically capture and embed global fuzzy style information from $z_a$ into a token $z_s \in \mathbb{R}^{1\times D'}$. This token is then broadcasted and combined with the universal audio latent representations $z_a \in \mathbb{R}^{T'\times D'}$ to form a unified latent representation $z_l \in \mathbb{R}^{T\times D''}$. We enhance the sequence's overall representational fidelity by merging local and global insights for co-speech gesture generation. Subsequently, the unified latent representation is directed to the downsampling module for further processing. 

The downsampling module is integrated into the condition extractor to ensure alignment between each latent representation and its corresponding sequence of encoded gestures. In our exploration, we experimented with linear alignment like DSG\cite{yang_diffusestylegesture_2023} and DSG+\cite{yang2023DiffuseStyleGestureaplus}, but noted an issue of foot-skating arising from these methods. On the contrary, We adopt a Conv1D layer with a kernel size of 201 for this module that maps every 201-length target token output from WavLM to one gesture frame. Finally, the fuzzy feature extractor outputs $c_{1:T}$, representing a unified latent representation that combines encoded audio features and diffusion time step $n$. The condition extractor can be formalized by:

\begin{align}\label{eq:conditionExtractor}
    &z_a = LE(a)     &&z_a \in \mathbb{R}^{T' \times D'}\notag\\ 
    &z_s = GE(z_a)      &&z_s \in \mathbb{R}^{1 \times D'}\notag \\
    &z_l = DS(z_a + z_s)&&z_l \in \mathbb{R}^{T \times D''}    \\
    &n' = DTE(n)        &&n \in \mathbb{R},~~ n' \in \mathbb{R}^{1 \times D''} \notag \\
    &c_{1:T} = z_l + n' &&c_{1:T} \in \ \mathbb{R}^{T \times D''} \notag
\end{align}
Where \(LE(\cdot)\) and \(GE(\cdot)\) denote the local extractor (WavLM) and the global extractor. \(DS(\cdot)\) represents the down sampling process. \(DTE(\cdot)\) signifies the diffusion time step embedding. The final output of the fuzzy feature extractor is denoted as \(c_{1:T}\). Here, \(T'\), \(D'\), and \(D''\) refer to the WavLM output token length, feature dimensionality of WavLM's output token, and the feature ($h$) dimensionality of the proposed model's hidden state, respectively. \(a\) is the input raw speech audio waveform. \(Z_a\) and \(Z_s\) are extracted by the local extractor(WavLM model) and the global extractor. \(Z_l\) is the unified latent representation. \(n\) is the diffusion time step, \(n'\) is the embedded diffusion time step feature. 

\subsubsection{AdaLN Transformer}
The AdaLN's fundamental purpose is to incorporate a conditional mechanism that uniformly applies a specific function across all tokens, thereby significantly improving the model's capacity for representing both conditional and output features with enhanced efficiency. It offers a more sophisticated and nuanced approach to modeling, enabling the system to capture and articulate the complex dynamics between various input conditions and their corresponding outputs. Consequently, this leads to an improvement in the model's predictive accuracy and its ability to generate outputs that are more aligned with the given conditions.  

Diffusion Transformers (DiTs)\cite{peebles2022Scalable} represent an advanced transformer-based backbone for diffusion models, surpassing previous U-Net models in performance. By incorporating AdaLN within transformer blocks for text-to-image synthesis, DiTs achieve lower Fréchet Inception Distance (FID)\cite{heusel2017gans} scores, indicating superior image quality. Recently, this framework has been used for text-conditional video generation. Despite Diffusion Transformers (DiTs) success in handling discrete text prompts conditional inputs, their effectiveness in speech-driven gesture generation, a sequence-to-sequence task, necessitates a thorough investigation.

Distinctively with the DiTs, our approach utilizes continuous fuzzy features as conditional input tokens. Further, it is without any patchy for spatial input, resulting in the output being the latent feature of a sequence of gestures. 

The module involves regressing the dimensionwise scale and shift parameters (\(\gamma\) and \(\beta\)), which are derived from the fuzzy feature extractor output \(c_{1:T}\), instead of directly learning \(\gamma\) and \(\beta\), as depicted in Figure \ref{fig:architecture3}.  In each AdaLN transformer, a latent feature denoted as \(z^n_{1:T,m}\) is generated by fusing condition information and gesture using AdaLN and causal self-attention. Here, \(1 \leq m \leq M\), where \(M\) represents the total number of AdaLN transformer stacks. In addition, the final layer, as illustrated in Figure \ref{fig:architecture2}, fed the same fuzzy features but with additional scale and shift operation.

This method facilitates the creation of detailed gesture sequences solely from speech audio, eliminating the requirement for discrete style labels or supplementary inputs. Consequently, it significantly improves the model's capacity to generate personalized and closely aligned gestures with the context of the speech, offering a more refined and context-sensitive gesture synthesis capability.

\subsubsection{Gesture Encoder and Decoder}
The architecture of the gesture encoder and decoder is designed to encode and decode the gesture sequence, as illustrated in Fig.\ref{fig:architecture1} and Fig.\ref{fig:architecture2}. The gesture encoder comprises a Convolution1D with a kernel size of 3. It encodes the initial sequence of gestures $g$ into a hidden state $h \in \mathbb{R}^{T \times D''}$. Our experimental results revealed that employing a kernel size of 1 resulted in animation jitter. Conversely, a kernel size of 3 is instrumental in mitigating this issue by effectively capturing the spatial-temporal relationships inherent in gesture sequences.  

The gesture decoder reduces the feature dimension of the output from the transformer \(D''\) to the original dimension \(D\), corresponding to the number of channels representing skeleton joints. Result in outputting the predicted noise (\(\epsilon_\theta\)). We utilize a size of 1 convolution1D By convolving a 1D kernel with each position in the input sequence, our model can effectively extract meaningful features and relationships between adjacent joint channels.
 % instead of employing a fully connected layer.

\subsection{Training and Inferencing with Denoising Diffusion Probabilistic Model} \label{sec:DDPM}
The diffusion process in this architecture aims to reconstruct the conditional probability distribution between gestures and fuzzy features. This entails employing a systematic approach to sample from this restored distribution, thereby enabling the generation of diverse gestures.

Following our previous work, Diffmotion\cite{zhang2023diffmotion}, incorporating the Denoising Diffusion Probabilistic Model (DDPM) into our approach. However, we employ a non-autoregressive transformer to generate the entire sequence of gestures instead of frame-by-frame. The form is represented by $p_\theta :=  \smallint p_\theta \left(g^{0:N}\right)dg^{1:N}$, where $g^1,...,g^N$ are latent of the same dimensionality as the data $g^n$ at the \(n\)-th diffusion time stage. 

The model contains two processes: the diffusion process and the generation process. At training time, the diffusion process gradually converts the original gesture data(\(g^0\)) to white noise(\(g^N\)) by optimizing a variational bound on the data likelihood. At inference time, the generation process recovers the data by reversing this noising process through the Markov chain using Langevin sampling\cite{paul_sur_1908}. The Markov chains in the diffusion process and the generation process are:
\begin{equation}
\begin{aligned}
&p\left(g^n|g^0\right) = \mathcal{N}\left(g^n; \sqrt{\overline{\alpha}^n} g^0, \left(1-\overline{\alpha}^n\right)I\right)   \quad and\\ 
&p_\theta\left(g^{n-1}|g^n, g^0\right) = \mathcal{N}\left(g^{n-1}; \tilde{\mu}^n\left(g^n, g^0\right), \tilde{\beta}^n I\right),
\end{aligned}
\label{eq:cumulativeProduct}
\end{equation}
where \(\alpha^n := 1 - \beta^n\) and \(\overline{\alpha}^n := \prod_{i=1}^n \alpha^i\). As shown by\cite{ho_denoising_2020}, \(\beta^n\) is a increasing variance schedule \(\beta^1,...,\beta^N\) with \(\beta^n \in (0,1)\), and \(\tilde{\beta}^n := \frac{1-\overline{\alpha}^{n-1}}{1-\overline{\alpha}^n}\beta^n\).

The training objective is to optimize the parameters \(\theta\) that minimizes the Negative Log-Likelihood (NLL) via Mean Squared Error (MSE) loss between the true noise \(\epsilon\sim\mathcal{N}\left(0,I\right)\)  and the predicted noise \(\epsilon_\theta\):
\begin{equation}
\label{eq:objective2}
\mathbb{E}_{g^0_{1:T}, \epsilon, n}[||\epsilon - \epsilon_\theta\left(\sqrt{\overline{\alpha}^n g^0}+\sqrt{1-\overline{\alpha}^n}\epsilon , a_{1:T},n\right)||^2],
\end{equation}  Here \(\epsilon_\theta\) is a neural network (see figure \ref{fig:architecture1}), which uses input \(g_t^0\) , \(a_{t-1}\) and \(n\) that to predict the \(\epsilon\), and contains the similar architecture employed in \cite{rasul_autoregressive_2021}. The complete training procedure is outlined in Algorithm \ref{alg:Training}.

\begin{algorithm}[htbp]
    \caption{Training for the whole sequence gesture}
    \KwIn{data $g^0_{1:T} \sim p\left(g^0|a_{1:T}\right)$ and \(a_{1:T}\)}
    \Repeat{converged}{Initialize $n \sim$ Uniform$(1,...,N)$ and $\epsilon \sim \mathcal{N}(0,I)$
    \\Take the gradient step on
    $$\nabla_\theta||\epsilon-\epsilon_\theta\left(\sqrt{\overline{\alpha}_n}g^0_{1:T}+\sqrt{1-\overline{\alpha}_n}\epsilon, a_{1:T},n\right)||^2$$}    \label{alg:Training}
\end{algorithm}

After training, we utilize variational inference to generate the whole sequence of new gestures matching the original data distribution(\(g^0 \sim p_\theta\left(g^0,a\right)\)). We followed the sampling procedure in Algorithm  \ref{alg:Inference} to obtain the whole sequence of the sample \(g^0\). The \(\sigma_\theta\) is the standard deviation of the \(p_\theta \left(g^{n-1}|g^n\right)\). We choose  \(\sigma_\theta := \tilde{\beta}^n\). 

\begin{algorithm}[htbp]
\SetKwFor{For}{for}{do}{end\enspace for}
\SetKwIF{If}{ElseIf}{Else}{if}{then}{else if}{else}{end\enspace if}
\SetKw{Return}{Return:}
\caption{Sampling $g_{1:T}^0$ via annealed Langevin dynamics}
\KwIn{ noise $g_{1:T}^N \sim \mathcal{N}(0,I)$ and raw audio waveform \(a_{1:T}\)} 
\For {$n = N$ \emph{\KwTo} $1$}{
	\eIf{$n>1$}{$z \sim \mathcal{N}(0,I)$}{$z = 0$}
	$g_{1:T}^{n-1}=\frac{1}{\sqrt{\alpha^n}}\left(g_{1:T}^n - \frac{\beta^n}{\sqrt{1-\overline{\alpha}^n}}\epsilon_\theta\left(g_{1:T}^n,a_{1:T},n\right)\right)+\sqrt{\sigma_\theta}z$
}
\Return{$g^0_{1:T}$}
\label{alg:Inference}
\end{algorithm}

During inferencing, we send the whole sequence of the raw audio to the condition extractor component. Then, the component output is fed to the Diffusion Model to generate the whole sequence of the accompanying gesture(\(g^0\)). 

\section{EXPERIMENTS}\label{sec:EXPERIMENTS}
To validate our approach, we utilized three co-speech gesture datasets (Trinity\cite{ferstlylva2018Investigating}, ZEGGS\cite{ghorbani2023ZeroEGGS}, and BEAT\cite{liu2022BEAT}). Our experiments concentrated on producing full 3D body gestures (including finger motions and locomotion). This choice presented a greater challenge than merely focusing on upper body motions due to the expanded output dimensionality and the need to overcome visual complexities, such as foot-skating, the naturalness of finger movements, and locomotion.

\subsection{Dataset and Data Processing}

\subsubsection{Datasets}
The Trinity dataset focuses on individual spontaneous speech, the ZEGGS dataset encompasses a wide range of emotional expressions, and the BEAT dataset consists of personalized movements exhibited by various individuals. 
% Further details are elaborated in Table\ref{tab:overview datasets} found in Appendix \ref{app_sec:datasets}. 

\subsubsection{Speech Audio Data Process}
In the Trinity dataset, the audio was recorded at a sampling rate of 44 kHz, while 48 kHz in ZEGGS and BEAT. However, due to the pre-training of the WavLM large model on speech audio sampled at 16 kHz, we uniformly resample all audio to match this frequency.

\subsubsection{Gesture Data Process}
We focus solely on full-body gestures, adopting the data processing techniques outlined by Alexanderson et al.\cite{alexanderson_simon_style-controllable_2020}. Given the variability in data quality and structure across motion datasets, we tailor our approach by selecting specific joints for analysis in each dataset. We omit hand skeleton data for the Trinity Gesture Dataset due to its inferior quality. For ZEGGS and BEAT datasets, our analysis includes finger joints and the same set of joints considered in the Trinity dataset. All data capture translational and rotational velocities to detail the root's trajectory and orientation. The datasets are uniformly downsampled to a frame rate of 20 fps. To ensure accurate and continuous representation of joint angles, we apply the exponential map technique\cite{grassiaf.sebastian1998Practical}. All data are segmented into 20-second clips for training and validation purposes. As for the user evaluation, we segment the generated gesture sequence into 10 seconds to improve the efficiency of the evaluation.

\subsection{Model Settings}
Our experiments employ 12 causal attention blocks, each comprising 16 attention heads (as depicted in Figure \ref{fig:architecture1}). The encoding process transforms each frame of the gesture sequence into hidden states \(h \in \mathbb{R}^{1280}\). For the WavLM model, we utilize the pre-trained WavLM Large model \footnote{https://github.com/microsoft/unilm/tree/master/wavlm}. To ensure temporal translation invariance, we employ a translation-invariant self-attention (TISA) mechanism\cite{wennberg2021Case}. 

The quaternary variance schedule of diffusion model starts from \(\beta_1 = 1 \times 10^{-4} \) till \(\beta_N = 5 \times 10^{-5}\) with linear beat schedule. The number of diffusion steps is \(N = 1000\). The training batch size is 32 per GPU.
% , with an AdamW optimizer and a learning rate of \(20 \times 10^{-5}\)

The model was developed using the Torch Lightning framework and tested on an Intel i9 processor with an A100 GPU. Training durations were approximately 4 hours for Trinity and ZeroEGGS and 21 hours for BEAT.

%%%%%%%%%%%%%%%%%%%%%%%%%%%%%%%%%%%%%%%%%%%%%%%%%%%%%%
\subsection{Visualization Results}
Our system excels in creating personalized gestures that align with the speech context, leveraging the fuzzy inference strategy to autonomously derive fuzzy features directly from speech audio. Furthermore, it showcases remarkable generalization and robustness by utilizing in-the-wild speech.

\begin{figure}[htbp]
    \centering
    \subfloat[Happy]{\includegraphics[width=0.24\textwidth]{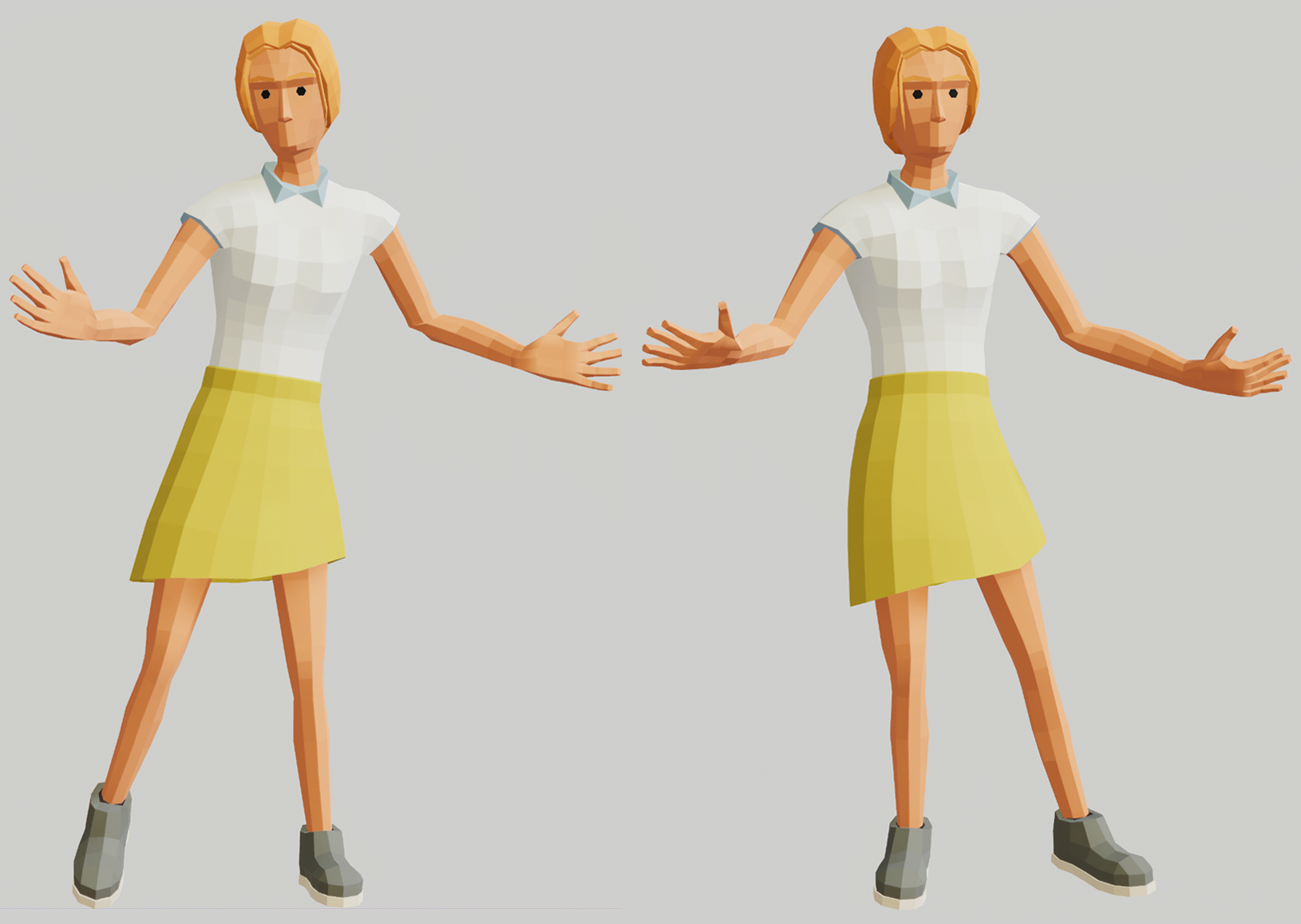}%
    \label{fig:happy} }
    \subfloat[Sad]{\includegraphics[width=0.24\textwidth]{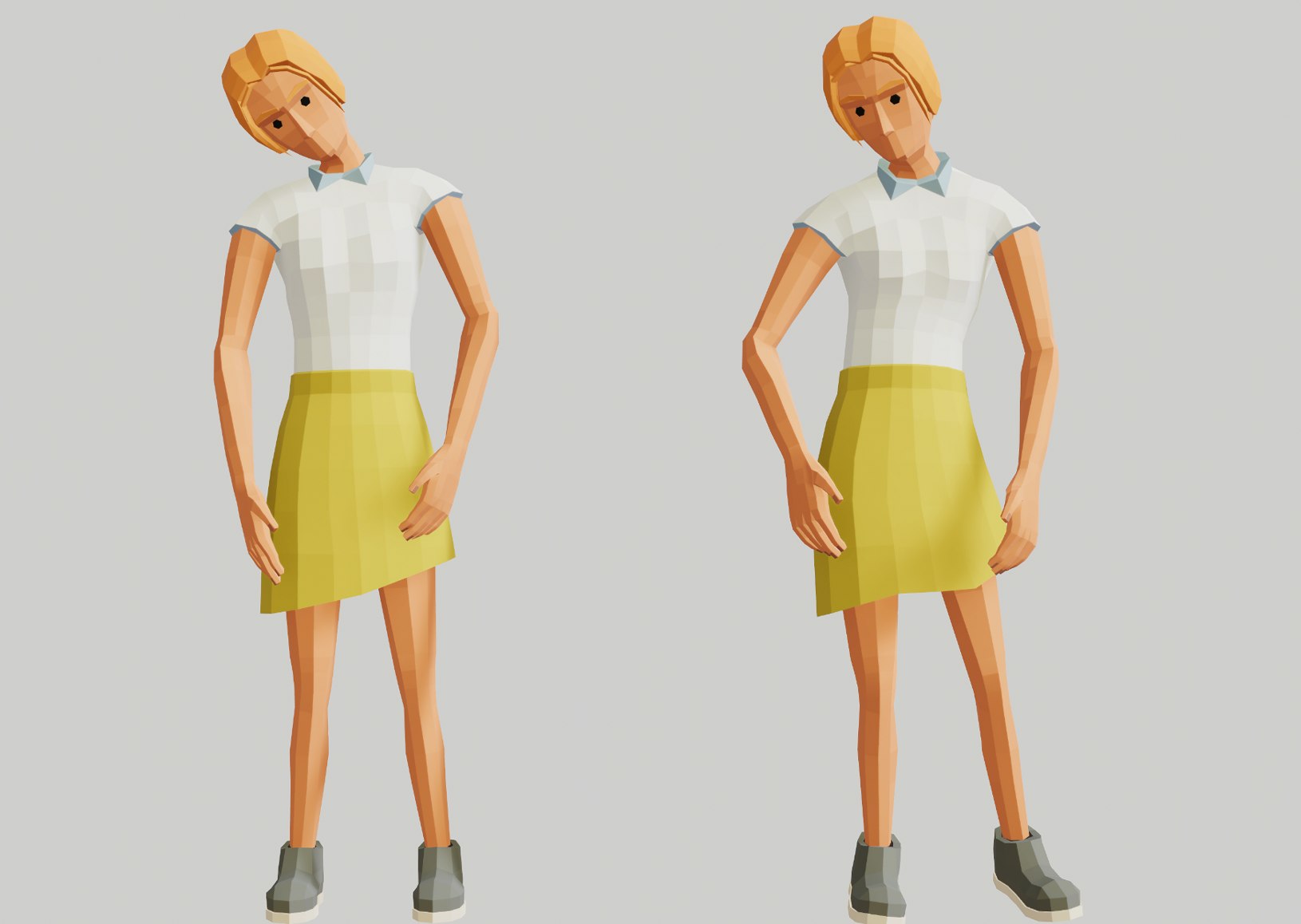}%
    \label{fig:Sad} }
    \hfil
    \subfloat[Old]{\includegraphics[width=0.24\textwidth]{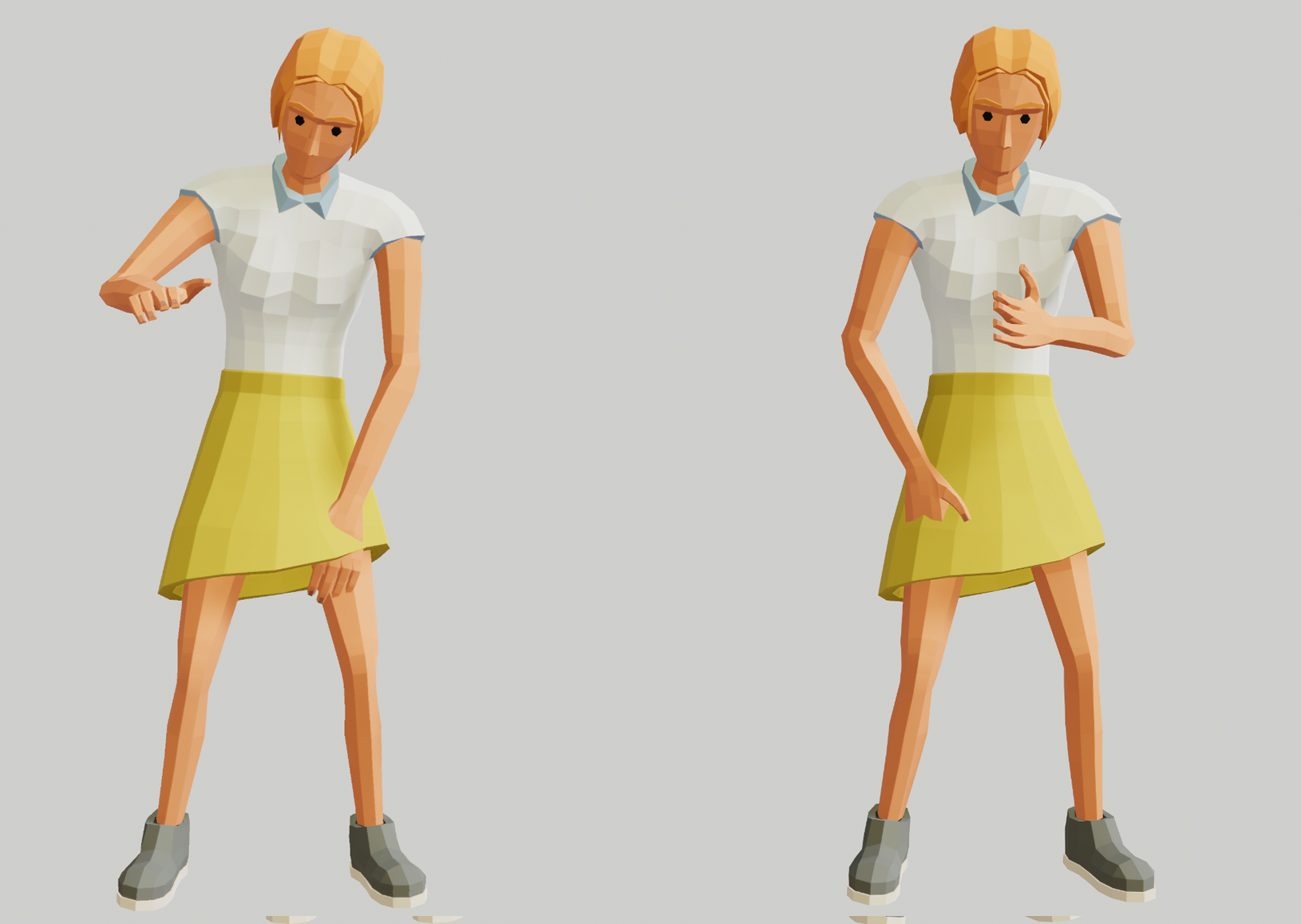}%
    \label{fig:Old} }
    \subfloat[Speech]{\includegraphics[width=0.24\textwidth]{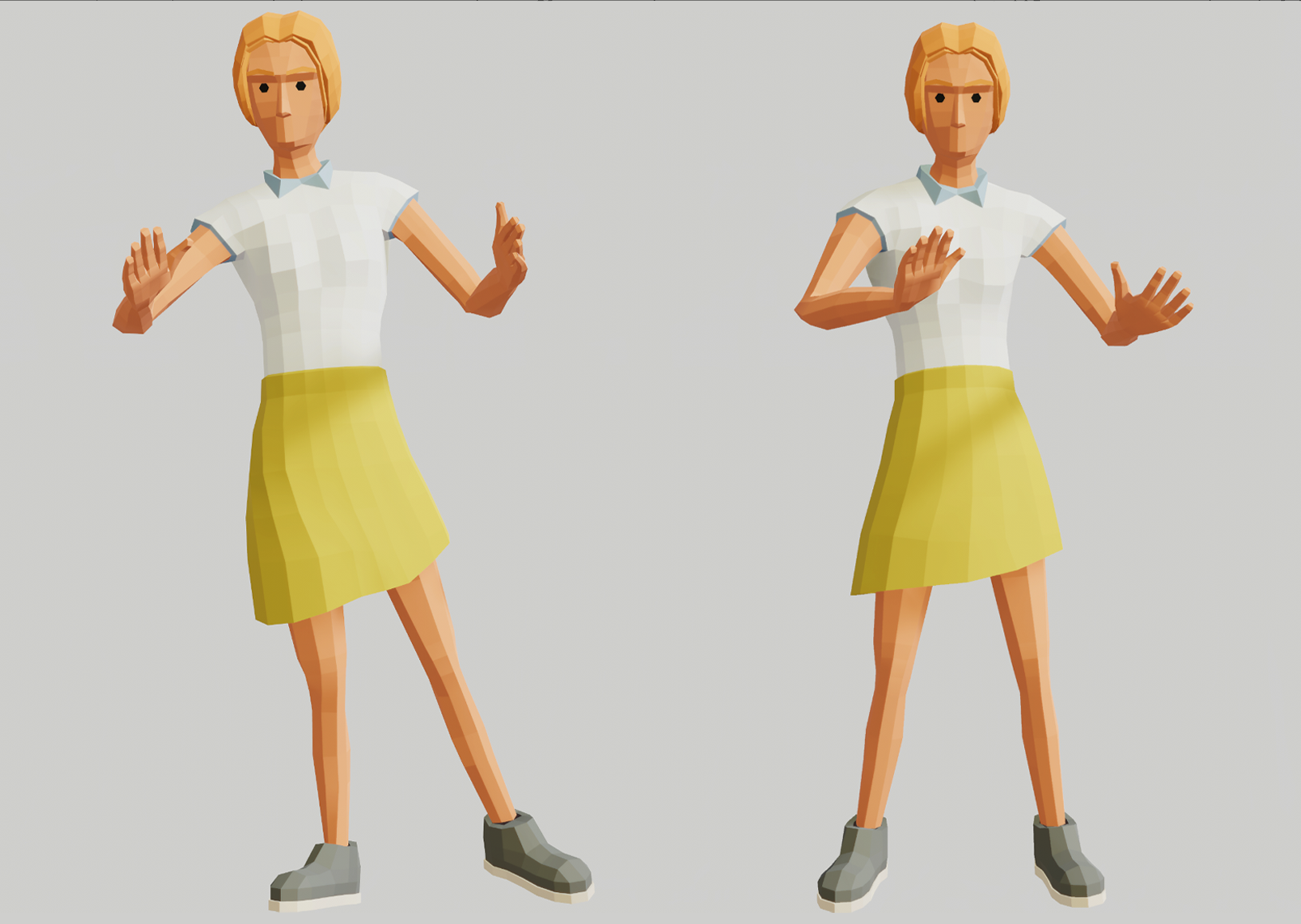}%
    \label{fig:Speech} }
    \caption{Samples of gestures corresponding to different emotions. The left side of the subfigure displays ground truth gestures, while the right side showcases gestures generated by our architecture.}
    \label{fig:Emotion}
\end{figure}

Figure \ref{fig:Emotion} depicts the visual outcomes of gestures aligned with the emotional valence conveyed by the audio. For instance, the system produces gestures of joy in response to happy audio cues (refer to Figure \ref{fig:happy}) and gestures of sadness for sorrowful audio (as depicted in Figure \ref{fig:Sad}). The system can also infer age-related characteristics or other nuanced states from the speech audio (as illustrated in \ref{fig:Old} and \ref{fig:Speech}). 

% \noindent
\begin{figure}[!htbp]
    \centering
    \subfloat[Ayana]{\includegraphics[width=0.24\textwidth]{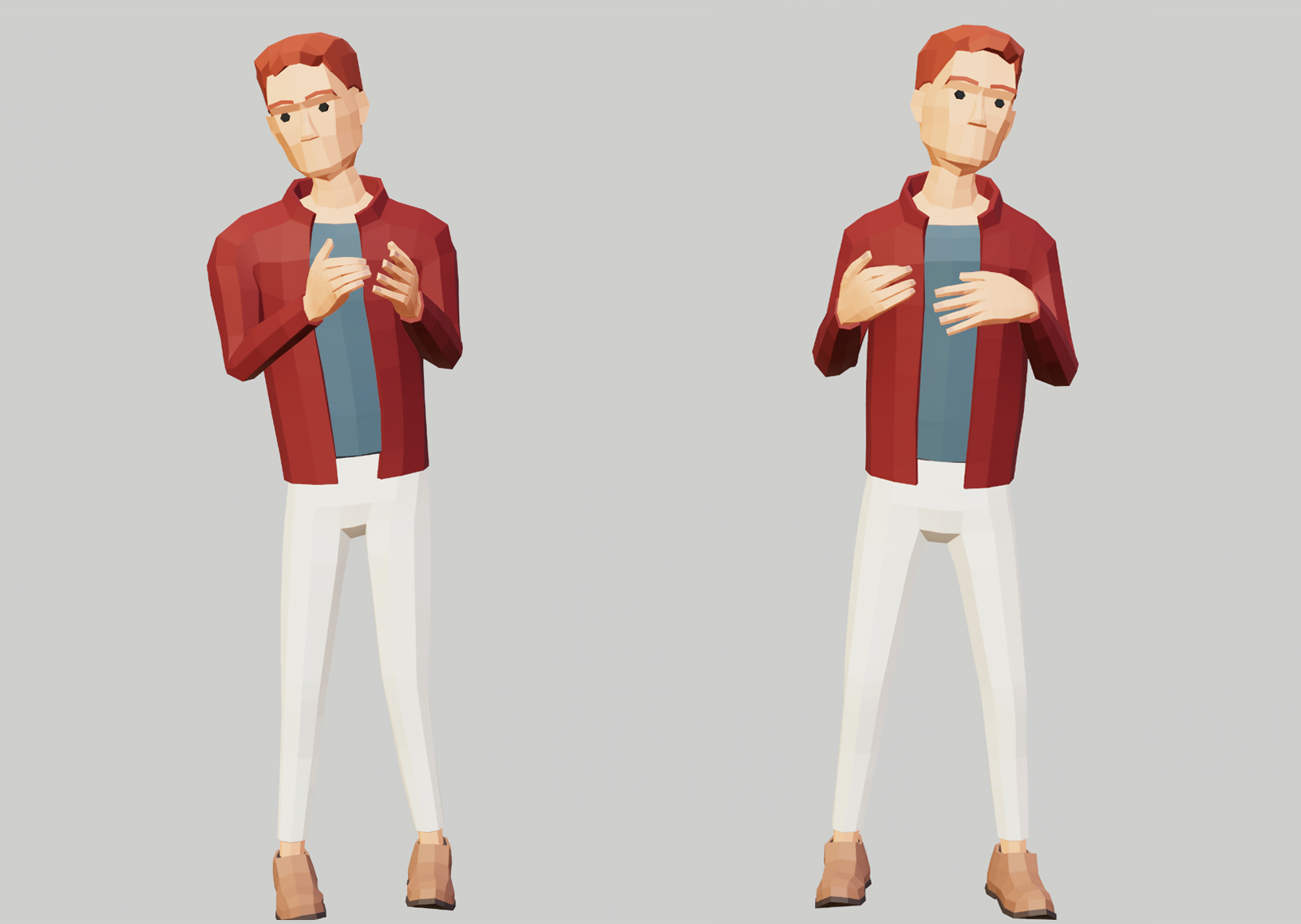}%
    \label{fig:Ayana} }
    \subfloat[Jaime]{\includegraphics[width=0.24\textwidth]{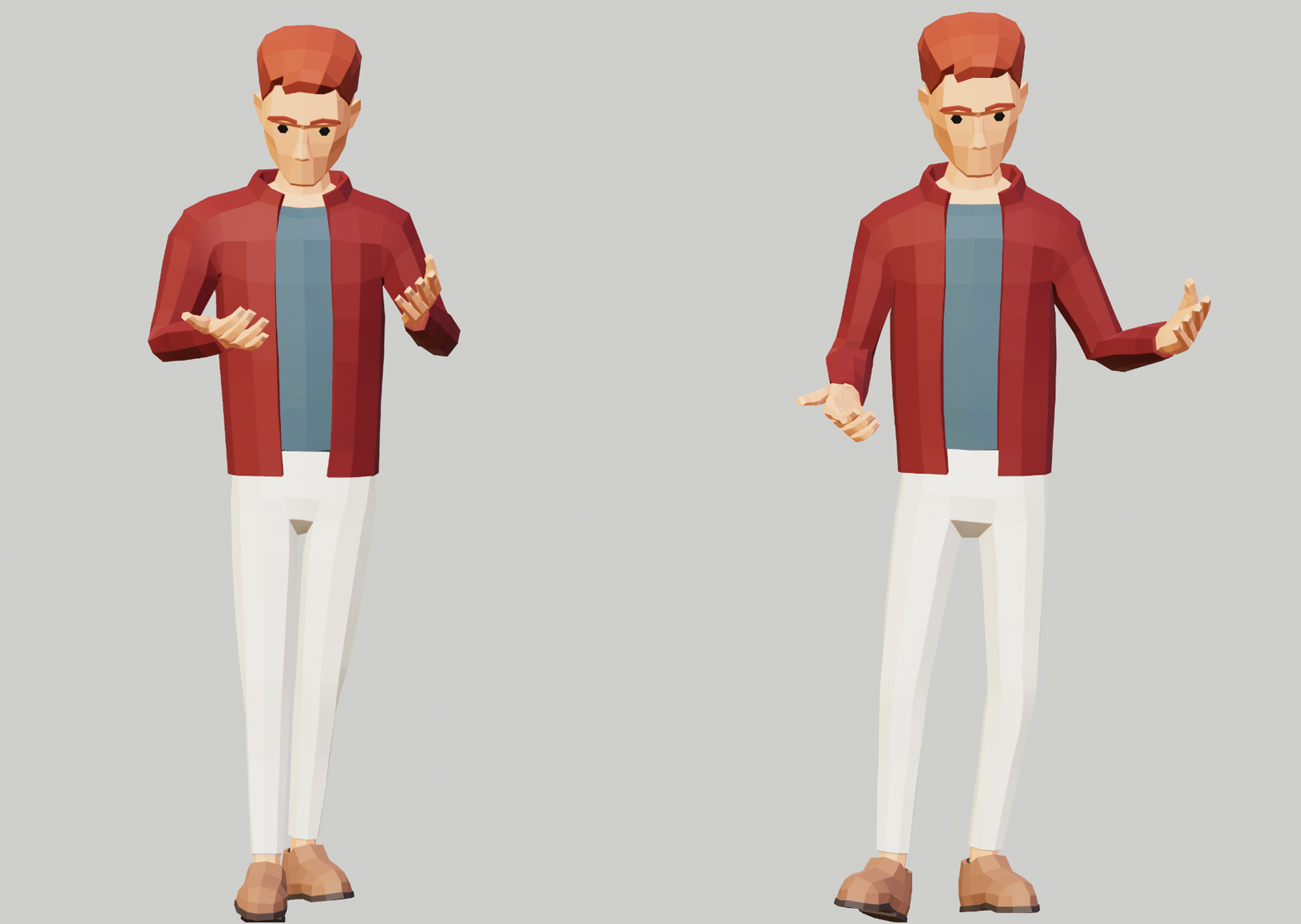}%
    \label{fig:Jaime} }
    \hfil
    \subfloat[Luqi]{\includegraphics[width=0.24\textwidth]{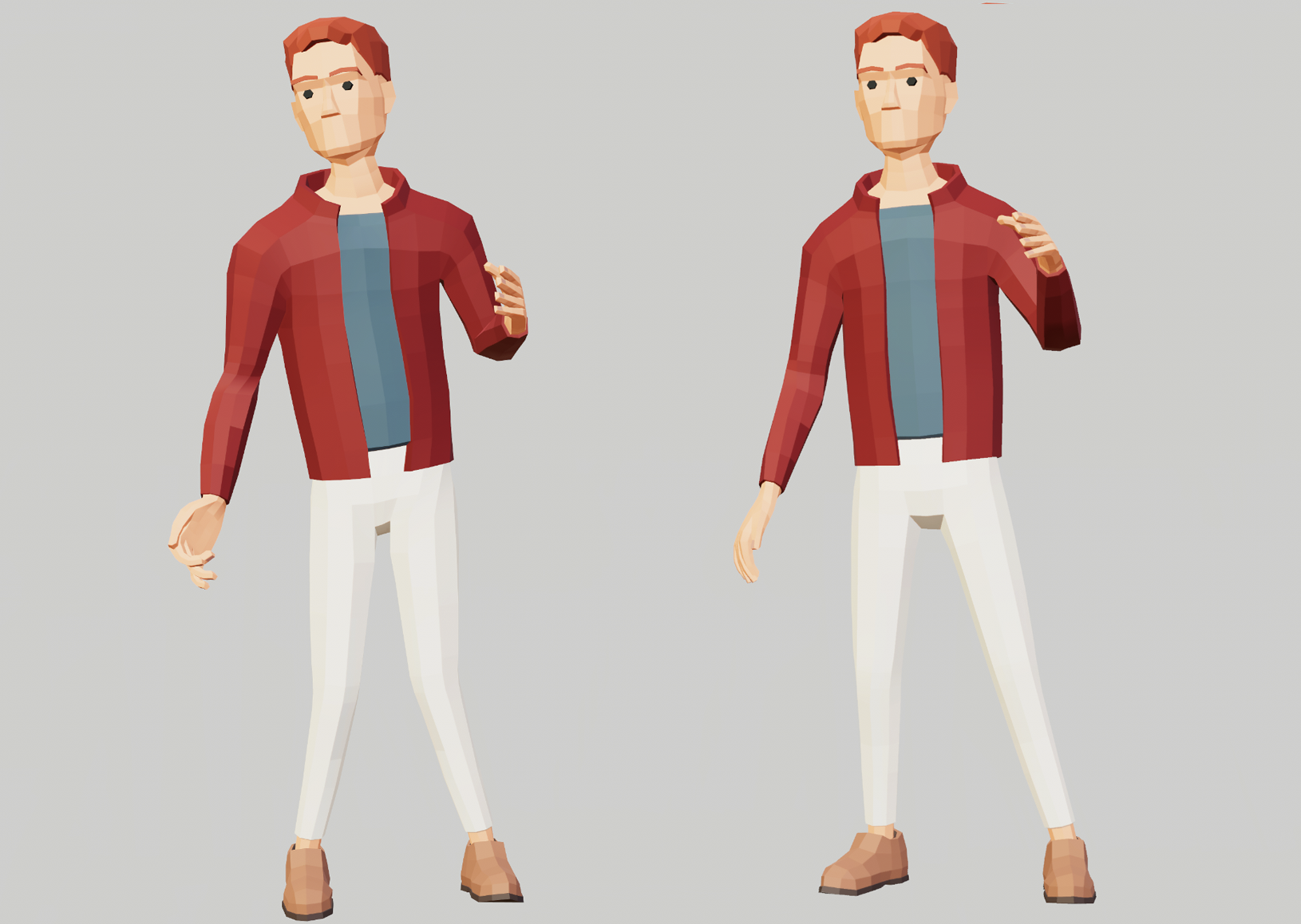}%
    \label{fig:Luqi} }
    \subfloat[Zhao]{\includegraphics[width=0.24\textwidth]{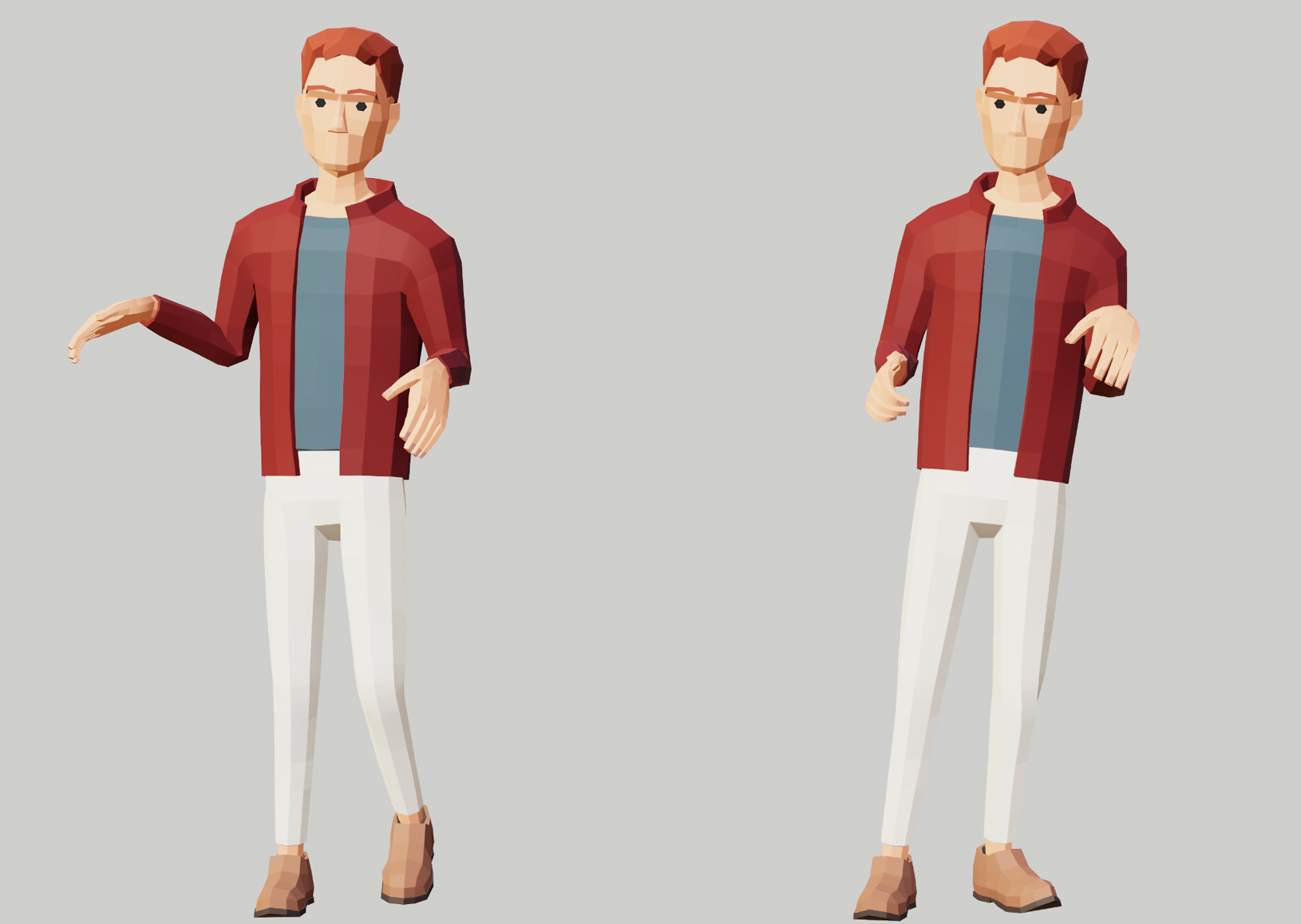}%
    \label{fig:Zhao} }
    \caption{Samples of gestures corresponding to different personalities. The left side of the subfigure displays ground truth gestures, while the right side showcases gestures generated by our architecture.}
    \label{fig:Personality}
\end{figure}

Figure \ref{fig:Personality} shows the system's ability to generate personalized gestures Predicated upon individuals' unique speech traits. For example, Ayana's gestures, with hands together and palms facing, denote reserved expressiveness. In contrast, Jaime's "palm up" gestures imply openness, and Luqi's alternating hand movements add dynamic variability. These results highlight the system's adeptness at depicting a wide range of personality-specific gestures.

% \noindent
\begin{figure}[!htbp]
    \centering
    \subfloat[...with \uline{my mother}...]{\includegraphics[width=0.16\textwidth]{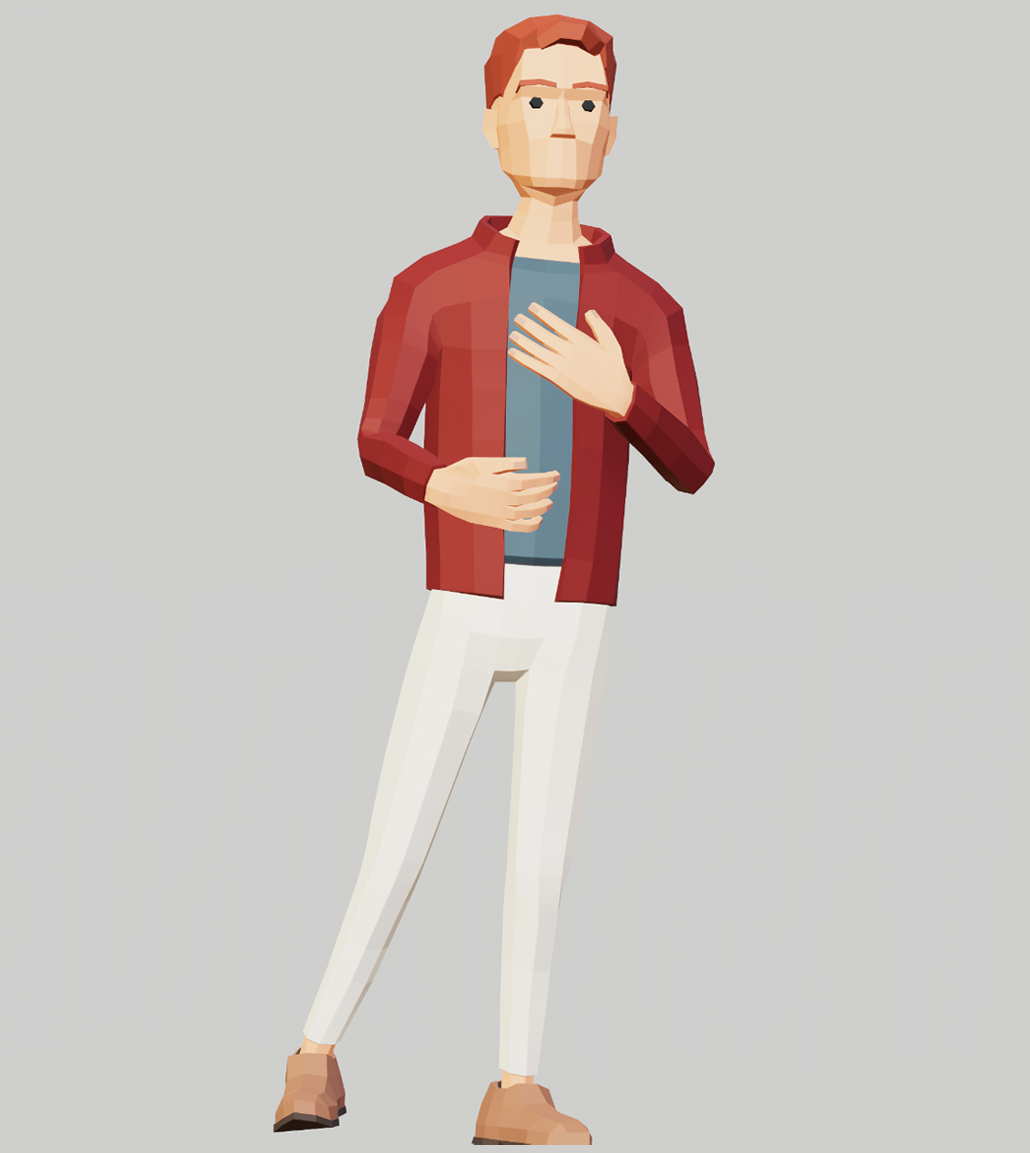}%
    \label{fig:carla_Sema} }
    \subfloat[...\uline{1400 miles away}...]{\includegraphics[width=0.16\textwidth]{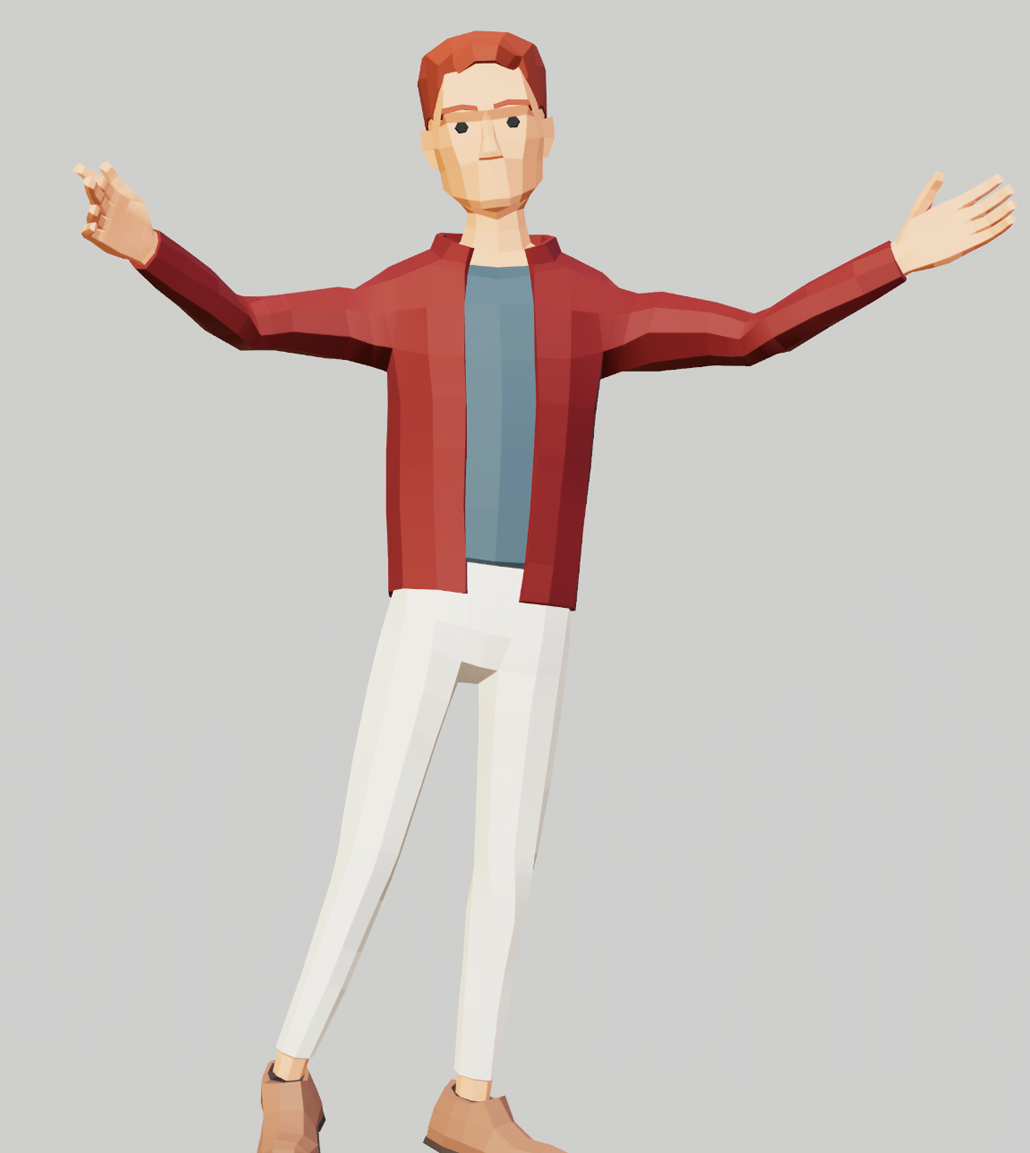}%
    \label{fig:lawrence_Sema} }
    \subfloat[...I\textsuperscript{'}m \uline{not saying}...]{\includegraphics[width=0.16\textwidth]{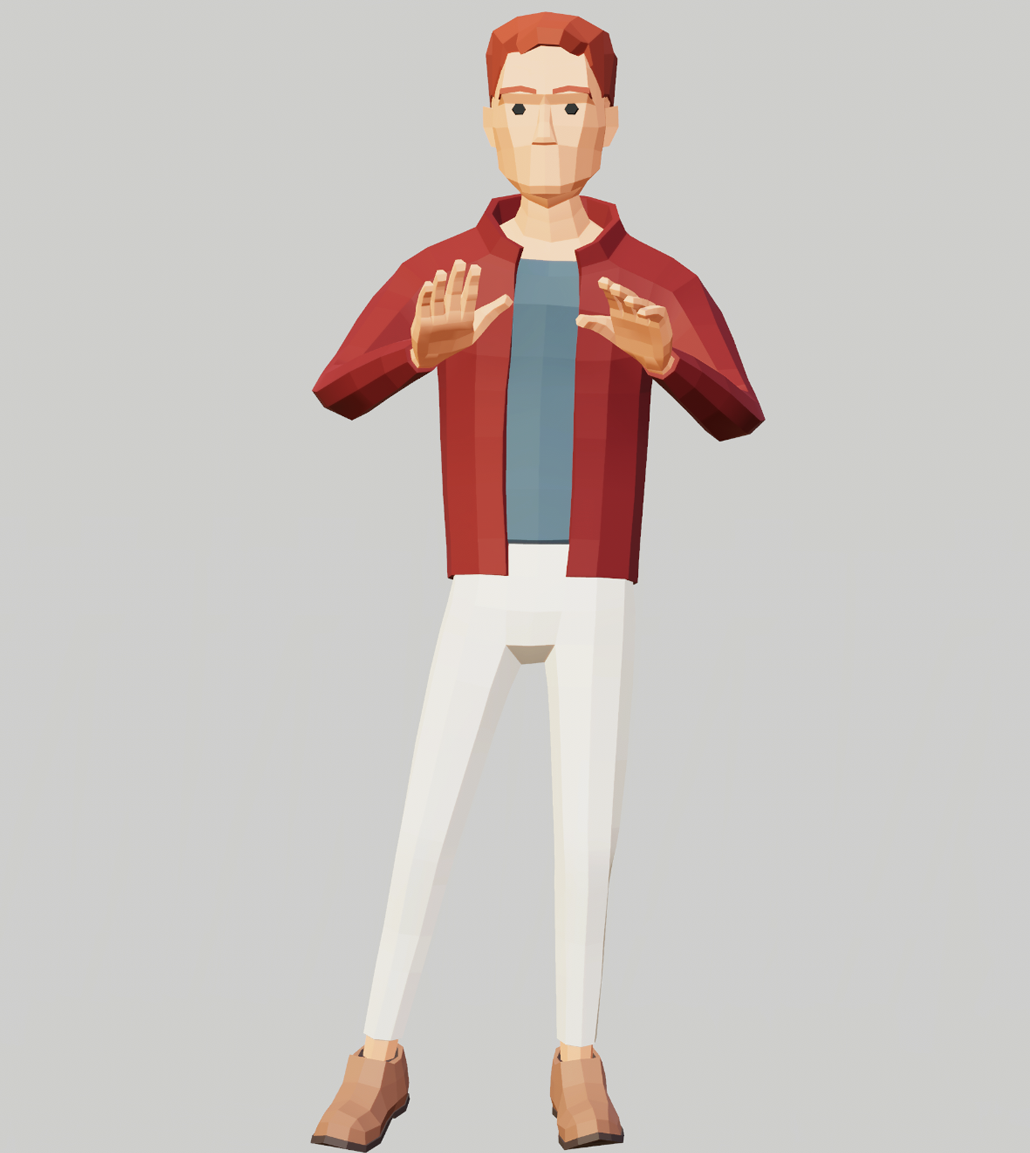}%
    \label{fig:threaten_Sema} }
    \caption{Samples of gestures corresponding to semantic.}
    \label{fig:semantic}
\end{figure}

% 替换图片，反映出动作的运动路径、手指等情况。
Interestingly, as shown in Figure \ref{fig:semantic}, the system can produce gestures with certain semantic relevance even in the absence of explicit semantic constraints. For instance, Carla's remark about her mother is matched with a self-referential gesture. Likewise, Lawrence's reference to distance is visually enhanced by a gesture that emphasizes the semantic essence of his speech.

% \noindent
\begin{figure}[htbp]
    \centering
    \includegraphics[width=0.48\textwidth]{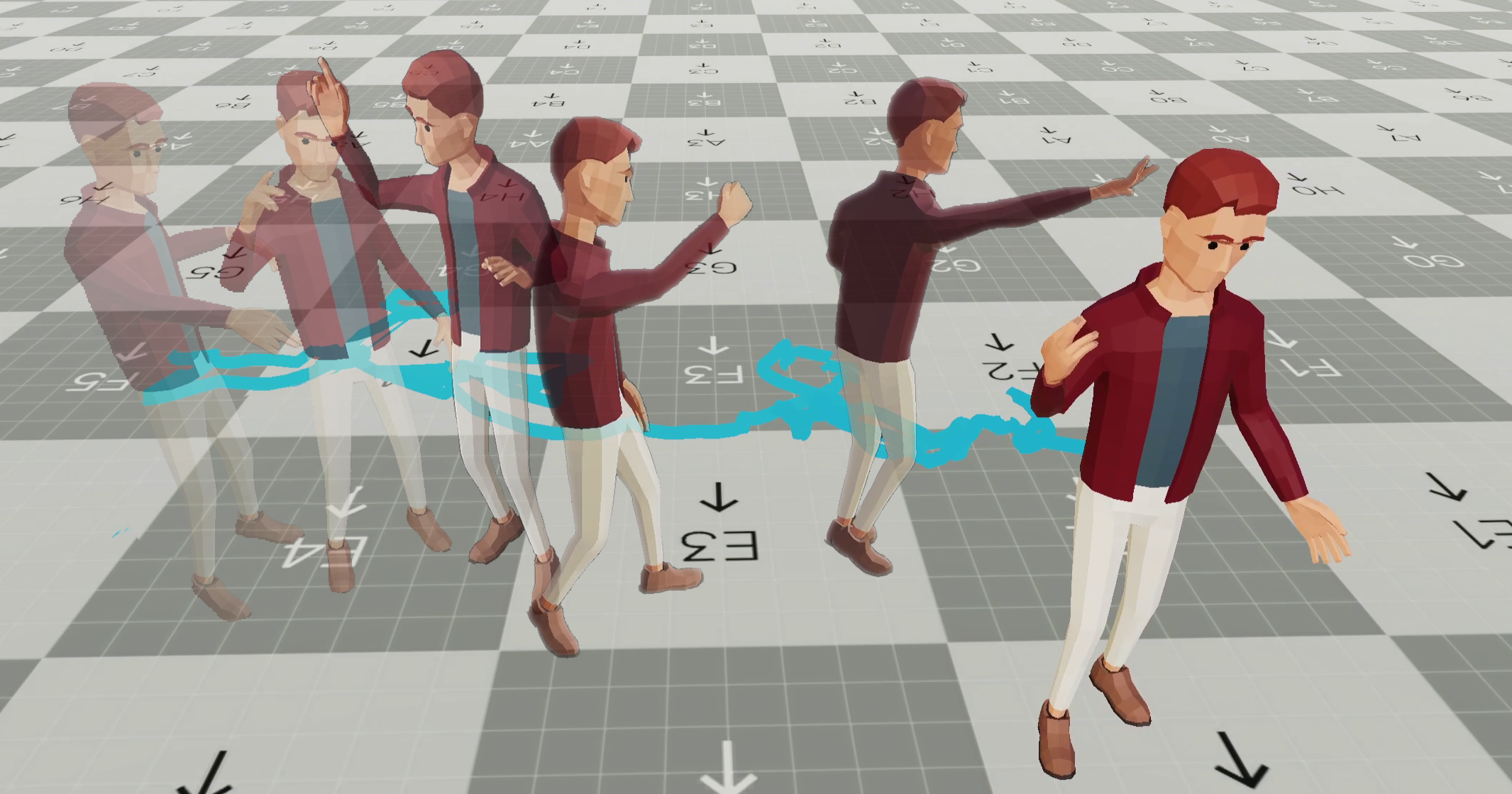}
    \caption{Sample of gesture including finger movements and locomotion.\protect\footnotemark}
    \label{fig:finger-loco}
\end{figure}

Further, finger movements and locomotion are included, as shown in Figure \ref{fig:finger-loco}, which highlight the system's proficiency in creating realistic, character-specific animations, thereby increasing the virtual interactions' believability and immersion.

\footnotetext{Due to the inherent challenges in retargeting finger motion to the avatar, please refer to the support video for more details.}

% \noindent
\begin{figure}[!htbp]
    \centering
    \subfloat[CamenAgraDeedy]{\includegraphics[width=0.16\textwidth]{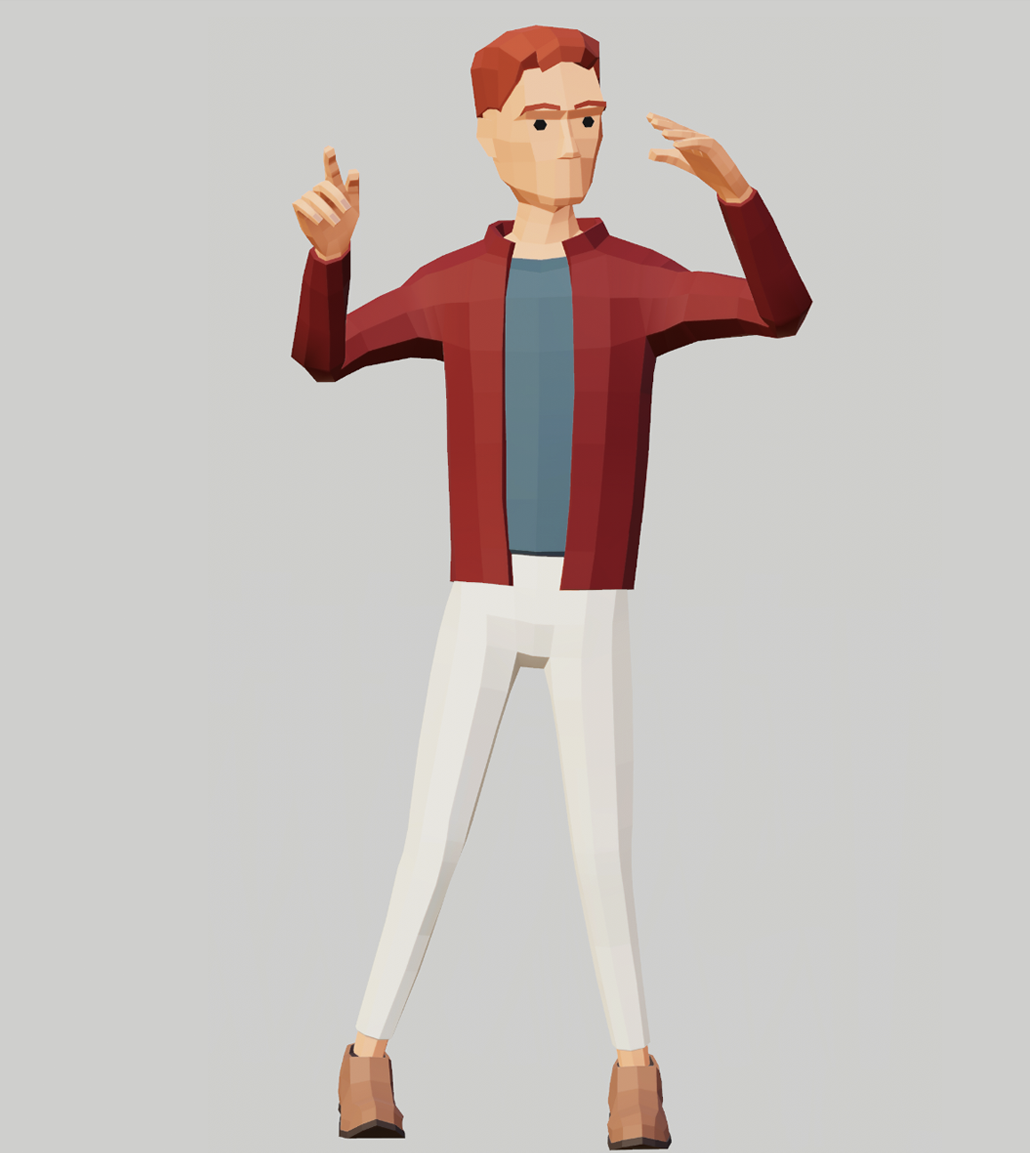}%
    \label{fig:CamenAgraDeedy} }
    % \hfil
    \subfloat[JinhaLee]{\includegraphics[width=0.16\textwidth]{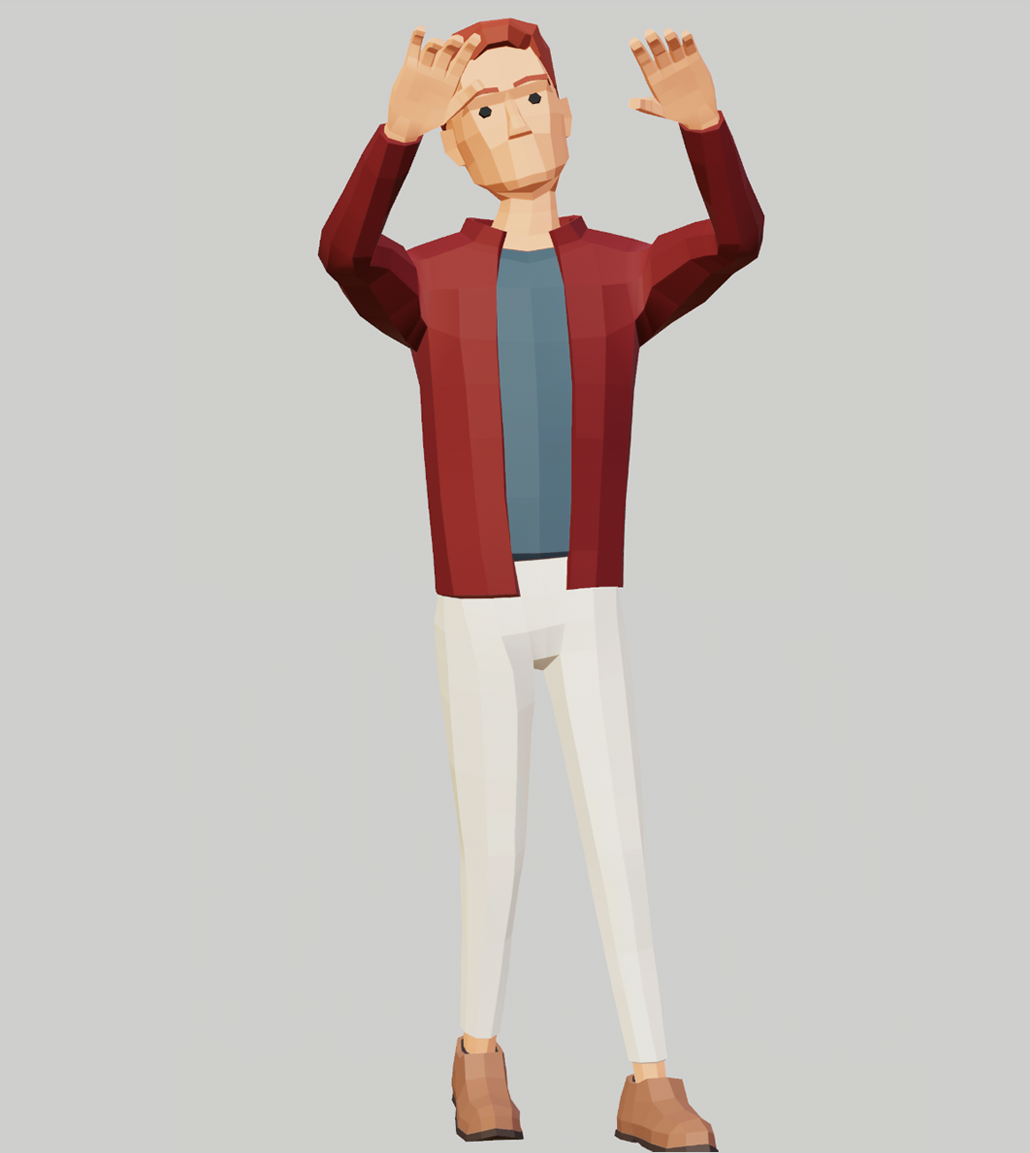}%
    \label{fig:JinhaLee} }
    % \hfil
    \subfloat[SakiMafunikwa]{\includegraphics[width=0.16\textwidth]{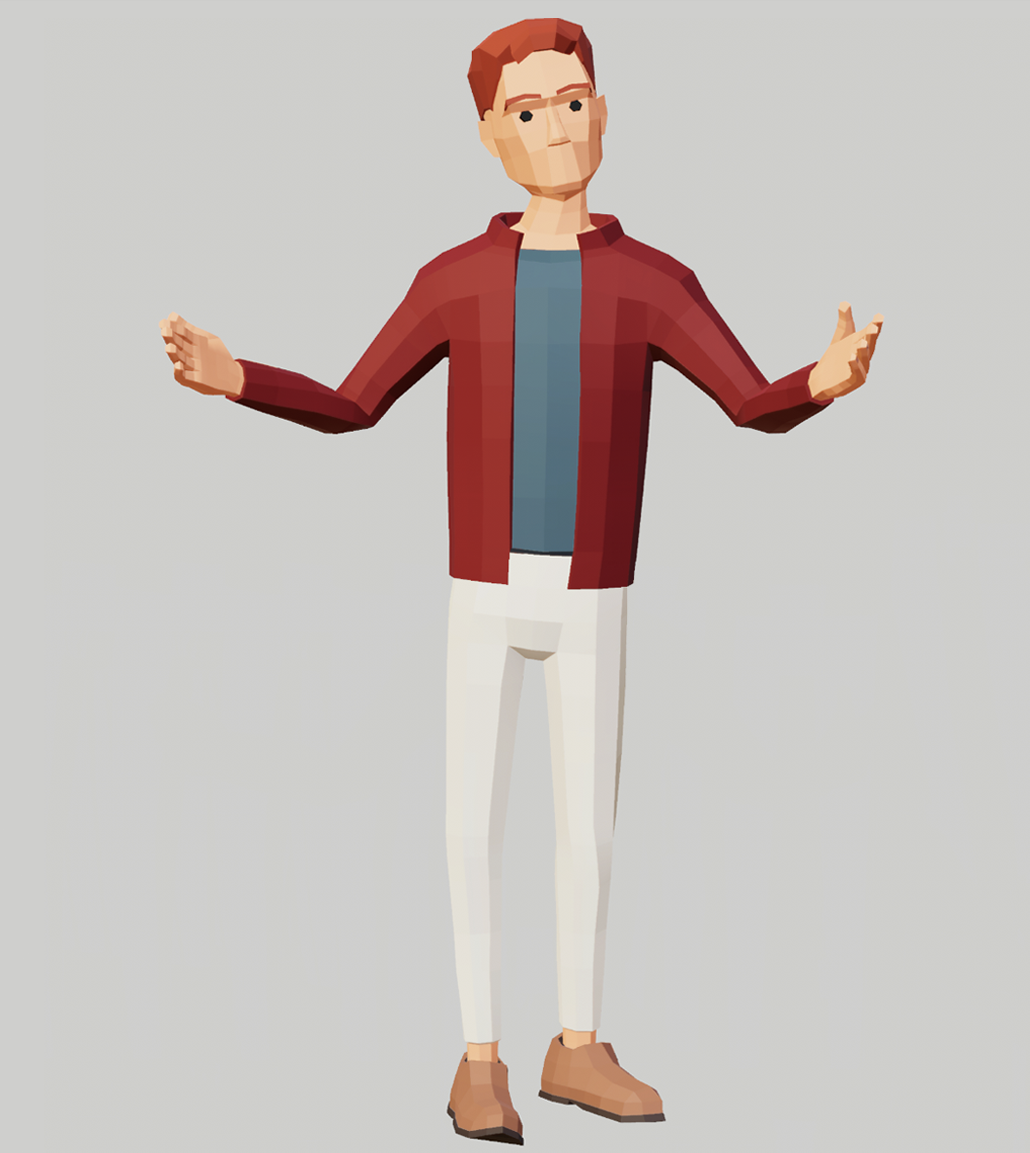}%
    \label{fig:SakiMafunikwa} }
    \caption{Samples of gestures corresponding to in-the-wild speech audio collected from TED Talks.}
    \label{fig:in_the_wild}
\end{figure}

The Figure \ref{fig:in_the_wild} showcase gesture outcomes generated from in-the-wild speech audio, like TED talks, to demonstrate the system's ability to create lifelike and style movements directly from unstructured real-world audio, without additional prompts or labels. This highlights the system's robust generalization capabilities. Testing in noisy environments with background music, applause, and urban sounds further revealed the system's strong anti-interference performance, emphasizing its resilience. This efficiency simplifies the input process, enabling effortless generation of dynamic character animations from raw audio, thus enhancing user experience and system accessibility. 

\begin{figure}[!htbp]
\centering
\subfloat[T-SNE results of the generated gestures in Zeggs Dataset experiment.]{\includegraphics[width=0.45\textwidth]{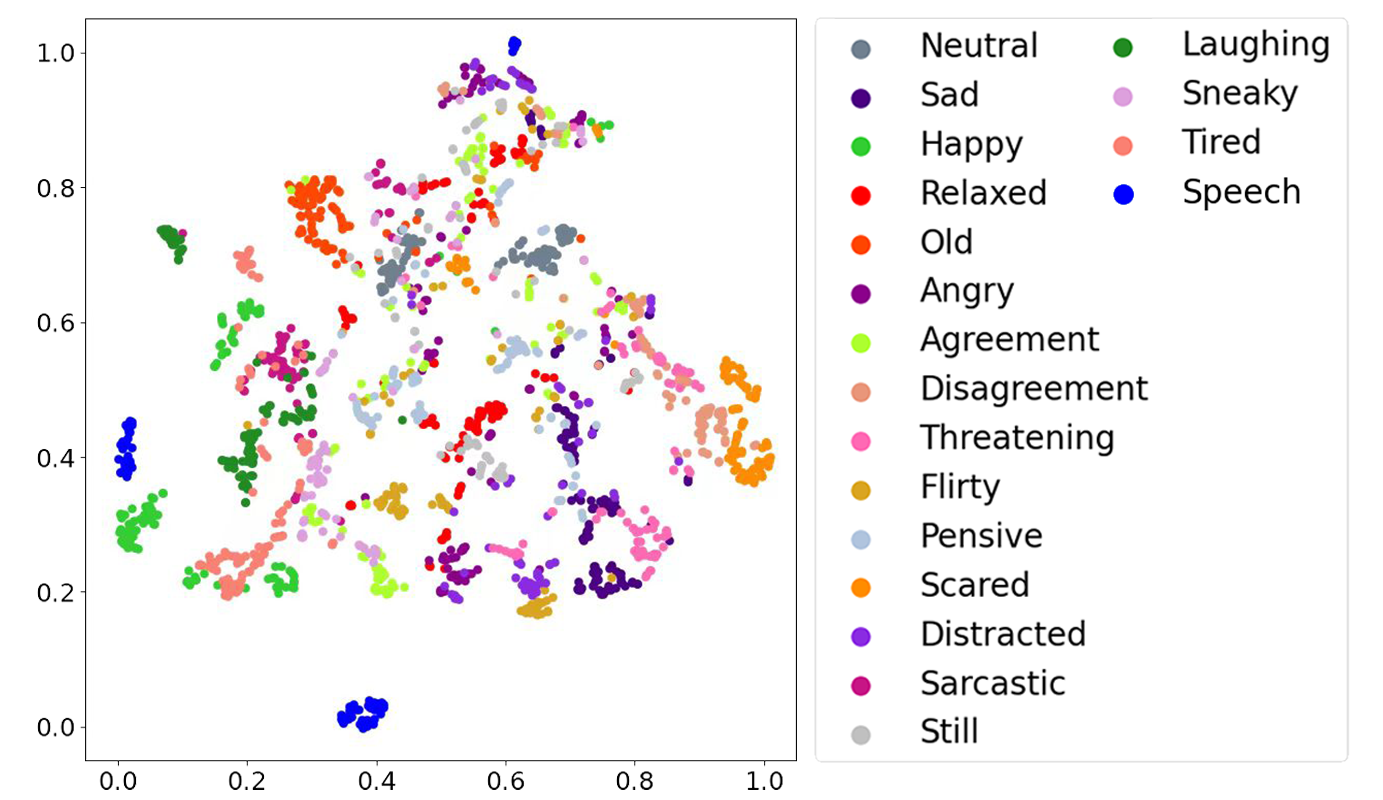}%
\label{fig:T-SNE-Zeggs} }
\hfil
\subfloat[T-SNE results of the generated gestures in BEAT Dataset experiment.]{\includegraphics[width=0.45\textwidth]{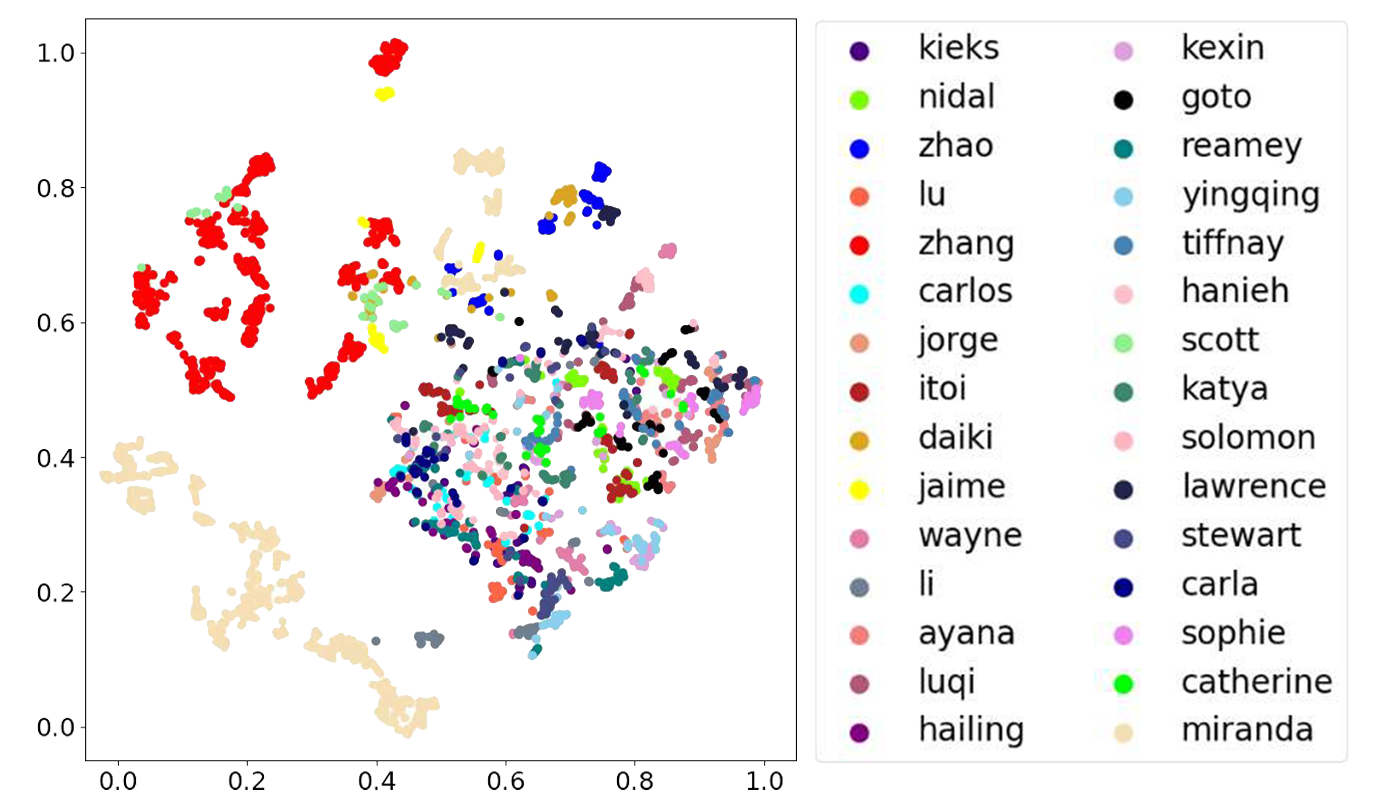}%
\label{fig:T-SNE-BEAT} }
\begin{center}
\caption{The T-SNE clustering visualization displays a variety of gestures distinguished by color-coded styles, revealing distinct regions for gestures tied to specific emotions or speakers, albeit with some boundary overlaps. This highlights our approach's capability to produce distinct style gestures through a fuzzy feature inference strategy, relying solely on speech audio.}
\label{fig:T-SNE}
\end{center}
\end{figure}

Finally, we represent the visualizes (Figure \ref{fig:T-SNE}) of the distribution of generated gestures corresponding to different emotional states (Fig. \ref{fig:T-SNE-Zeggs}) and personalities(Fig. \ref{fig:T-SNE-BEAT}) using the t-SNE method. The figure illustrates distinct separations between certain states, while others exhibit a degree of similarity yet remain distinguishable. These findings demonstrate the capability of our proposed method to generate nuanced and discernible gestures solely from raw speech audio without relying on labels or manual annotations. 

% \noindent

\subsection{Subjective and Objective Evaluation}
Consistent with the prevailing practices in gesture generation research, we conducted a series of subjective and objective evaluations to evaluate the co-speech gestures generated by our proposed Persona-Gestor (PG) model.
% \newsavebox{\mycitation} % 定义一个新的盒子
% \savebox{\mycitation}{\cite{alexanderson2023listen}} % 将 \cite 的结果保存到盒子中

We adopted slightly varied baselines for different datasets. For the Trinity dataset, we employed LDA\cite{alexanderson2023listen} and Taming\cite{zhu_taming_2023}. In addition to LDA and Taming, for the ZEGGS dataset, we also incorporated DiffuseStyleGesture (DSG)\cite{yang_diffusestylegesture_2023} and ZeroEGGS\cite{ghorbani2023ZeroEGGS} Furthermore, for the BEAT dataset, we utilized the same baseline models as in ZEGGS but replaced DSG with DSG+\cite{yang2023DiffuseStyleGestureaplus} and introduced GestureDiffuCLIP (GDC)\cite{ao2023GestureDiffuCLIP} as an additional baseline model. 
% \footnote{Despite the model not being predicated on diffusion and Transformer architectures, it exhibits exceptional performance and achieves the best performance among all deep-based models in the 2022 GENEA.}.

In our experiments with the ZEGGS and BEAT datasets, we extended the original LDA, DSG, and DSG+\footnote{The authors have expanded their coverage to include all types in the BEAT dataset, as originally presented in the project of that study.} models to cover all styles within these datasets. Originally, the Taming model, trained exclusively on the TED dataset, focused on upper-body gestures. We have since augmented it to support full-body gestures across the three datasets. 
% Efforts to adapt LDA to include finger motions were met with challenges, leading to unsatisfactory outcomes in gesture generation. Consequently, we utilized LDA-generated gestures, excluding finger movements, for our analysis. 
% For more implementation details of these baselines, please refer to Appendix \ref{app:issueofLADandDSG}.

\begin{table*}[h]
\caption{The subject mean perceptual rating score. Bold fonts were utilized to emphasize the best results for each metric among the different methods, except for the GT.}
\label{tab:overview}
\centering
\resizebox{\linewidth}{!}{
\begin{tabular}{ccccccccc}
\hline
                         & Methods        &                                                        & \multicolumn{3}{c}{Subject Evaluation Metric}                                                                                                                                                   & \multicolumn{3}{c}{Objective Evaluation Metric}                                                                                                    \\ \hline
Dataset                  & Model          & With Fingers & \begin{tabular}[c]{@{}c@{}}Human↑\\ likeness\end{tabular} & \begin{tabular}[c]{@{}c@{}}appropriateness↑\end{tabular} & \begin{tabular}[c]{@{}c@{}}Style↑\\ appropriateness\end{tabular} & \begin{tabular}[c]{@{}c@{}}FGD↓\\ on feature space\end{tabular} & \begin{tabular}[c]{@{}c@{}}FGD↓\\ on raw data space\end{tabular} & BeatAlign↑    \\ \hline
\multirow{8}{*}{Trinity} & GT             & \textit{Y}                                                      & 0.51±1.73                                                 & 0.66±1.24                                                       & /                                                                & /                                                               & /                                                                & /             \\
                         & LDA\cite{alexanderson2023listen}            & \textit{N}                                                      & -0.22±0.98                                                & -0.39±1.08                                                      & -0.18±1.10                                                      & 349.53                                                          & 6008.37                                                          & 0.68          \\
                         & Taming\cite{zhu_taming_2023}         & \textit{N}                                                      & -0.48±0.96                                                & -0.47±1.07                                                      & -0.23±1.05                                                      & 3970.14                                                         & 52196.87                                                         & 0.68          \\
                         & (Proposed)PG   & N                                                      & \textbf{0.12±1.09}                                        & \textbf{0.19±1.12}                                              & \textbf{0.20±1.06}                                              & \textbf{289.42}                                                 & \textbf{5080.57}                                                 & \textbf{0.69} \\
                         & (Ours)PGNSE    & N                                                      & 0.08±1.06                                                 & 0.13±1.18                                                       & 0.15±1.03                                                       & 8002.65                                                         & 70617.28                                                         & 0.68          \\
                         & (Ours)PGCA     & N                                                      & -0.17±1.02                                                & -0.01±1.11                                                      & -0.03±1.02                                                      & 2566.65                                                         & 26124.39                                                         & 0.68          \\
                         & (Ours)PGCF     & N                                                      & -0.08±1.06                                                & -0.13±1.05                                                      & 0±1.02                                                           & 12540.17                                                        & 156895.12                                                        & 0.68          \\
                         & (Ours)PGICC    & N                                                      & 0.07±1.01                                                 & 0.06±1.12                                                       & 0.08±1.01                                                       & 125921.53                                                       & 1940124.44                                                       & 0.67          \\ \hline
\multirow{11}{*}{ZEGGS}  & GT             & Y                                                      & 0.95±1.13                                                 & 1.19±1.03                                                        & /                                                                & /                                                               & /                                                                & /             \\
                         & LDA\cite{alexanderson2023listen}            & N                                                      & -0.73±1.12                                                & 0.02±1.31                                                       & 0.53±1.41                                                        & 124.55                                                          & 50996.33                                                         & 0.66          \\
                         & DSG\cite{yang_diffusestylegesture_2023}            & Y                                                      & -0.47±1.14                                                & -0.71±1.11                                                      & -0.61±1.08                                                       & 66.77                                                           & 33297.50                                                         & 0.63          \\
                         & Taming\cite{zhu_taming_2023}         & Y                                                      & -1.08±1.01                                                & -0.91±1.09                                                      & -0.98±0.97                                                       & 1419.76                                                         & 293245.12                                                        & 0.67          \\
                         & ZeroEGGS\cite{ghorbani2023ZeroEGGS}       & Y                                                      & 0.38±1.11                                                 & 0.29±1.35                                                        & 0.49±1.31                                                        & 37.19                                                           & 26666.85                                                         & 0.66          \\
                         & (Proposed)PG   & Y                                                      & \textbf{0.42±1.17}                                        & \textbf{0.48±1.29}                                              & \textbf{0.76±1.34}                                               & \textbf{28.13}                                                  & \textbf{26193.92}                                                & \textbf{0.68} \\
                         & (Ours)PGNSE    & Y                                                      & 0.33±1.15                                                 & 0.35±1.29                                                       & 0.59±1.38                                                        & 125.40                                                          & 49081.55                                                         & 0.66          \\
                         & (Ours)PGOnehot & Y                                                      & 0.25±1.19                                                 & 0.33±1.31                                                       & 0.51±1.28                                                        & 122.56                                                          & 50259.95                                                         & 0.63          \\
                         & (Ours)PGCA     & Y                                                      & -0.36±1.22                                                & -0.62±1.14                                                      & -0.66±1.09                                                       & 807.12                                                          & 156686.01                                                        & 0.67          \\
                         & (Ours)PGCF     & Y                                                      & 0.27±1.20                                                 & -0.03±1.28                                                      & -0.23±1.26                                                       & 97.57                                                           & 39256.55                                                         & 0.67          \\
                         & (Ours)PGICC    & Y                                                      & 0.04±1.20                                                 & -0.39±1.23                                                      & -0.40±1.21                                                       & 407.89                                                          & 89893.96                                                         & 0.67          \\ \hline
\multirow{11}{*}{BEAT}   & GT             & Y                                                      & 0.65±1.16                                                 & 0.96±1.04                                                        & /                                                                & /                                                               & /                                                                & /             \\
                         & LDA\cite{alexanderson2023listen}            & \textit{N}                                                      & -1.65±0.73                                                & -1.59±0.74                                                       & -1.35±1.05                                                       & 276.25& 3584.95& 0.66          \\
                         & DSG+\cite{yang2023DiffuseStyleGestureaplus}            & Y                                                      & -0.28±1.17                                                & -0.49±1.15                                                       & -0.40±1.24                                                       & 23811.46                                                        & 2384465.64                                                       & 0.43          \\
                         & GDC\cite{ao2023GestureDiffuCLIP}            & \textit{N}                                                      & 0.54±1.12                                                 & 0.47±1.25                                                        & 0.30±1.27                                                        & 432.15                                                          & 93215.56                                                         & \textbf{0.69} \\
                         & Taming\cite{zhu_taming_2023}         & Y                                                      & -0.42±1.14                                                & -0.52±1.14                                                       & -0.32±1.24                                                       & 1251.56                                                         & 46828.23                                                         & 0.66          \\
                         & (Proposed)PG   & Y                                                      & \textbf{0.56±1.14}                                        & \textbf{0.63±1.10}                                               & \textbf{0.66±1.16}                                               & \textbf{264.06}& \textbf{3471.26}& 0.68          \\
                         & (Ours)PGNSE    & Y                                                      & 0.09±1.16                                                 & 0.27±1.23                                                        & 0.46±1.31                                                        & 1514.94                                                         & 51077.98                                                         & 0.66          \\
                         & (Ours)PGOnehot & Y                                                      & -0.01±1.16                                                & 0.18±1.31                                                        & 0.32±1.36                                                        & 1863.69                                                         & 63872.78                                                         & 0.63          \\
                         & (Ours)PGCA     & Y                                                      & 0.35±1.09                                                 & 0.17±1.26                                                        & 0.28±1.33                                                        & 703.83                                                          & 18990.56                                                         & 0.66          \\
                         & (Ours)PGCF     & Y                                                      & 0.14±1.01                                                 & 0.15±1.22                                                        & 0.30±1.31                                                        & 1160.63                                                         & 48899.63                                                         & 0.66          \\
                         & (Ours)PGICC    & Y                                                      & 0.02±1.10                                                 & -0.24±1.17                                                       & -0.25±1.25                                                       & 2057.31                                                         & 78754.92                                                         & 0.66          \\ \hline
\end{tabular}
}
\end{table*}

\subsubsection{Subjective Evaluation}

The goal of speech-driven gesture generation is to produce gestures that are both natural and convincing. However, exclusive reliance on objective metrics may not adequately reflect human subjective quality assessments\cite{alexanderson_simon_style-controllable_2020, wolfertReviewEvaluationPractices2022, kucherenko2023Evaluating}. This study prioritizes subjective evaluations to gauge human perception, complemented by objective evaluations detailed in Section \ref{sec:objective eval}.
% \usepackage{multirow}

% \usepackage{multirow}

For thorough subjective evaluations, we utilize three metrics: human likeness, appropriateness, and style appropriateness. Human likeness gauges the naturalness and resemblance of gestures to real human movements independent of speech. Appropriateness examines the temporal alignment of gestures with speech rhythm, intonation, and semantics, ensuring natural fluidity. Style-appropriateness evaluates the similarity between generated and original gestures.

We conducted a user study with pairwise comparisons, as recommended by\cite{wolfert_rate_2021}. In each trial, participants were shown two 10-second video clips generated by different models (including the Ground Truth (GT)) side by side for direct comparison. The videos were accompanied by instructions for participants to select their preferred clip based on their evaluations. Preferences were quantified on a 0 to 2 scale, with the unselected clip in each pair receiving an inverse score (e.g., a -2 score for the non-chosen clip if the chosen one received 2). A score of zero indicated no preference. Attention checks were included in the study to ensure engagement. 
% Further details are available in Appendix \ref{appendix: userStudy}.

Considering the extensive range of styles in ZEGGS (19) and BEAT (30), individual evaluations for each style were deemed impractical. Consequently, we utilized a random selection method to assign a subset of 5 styles from ZEGGS and 6 characters from BEAT to each participant. For the Trinity dataset, we chose Record\textunderscore008 and Record\textunderscore015. The training or validation sets include none of the selected audio clips.
% \noindent
\begin{figure*}[!htbp]
    
    \centering
    \subfloat[Trinity]{\includegraphics[width=0.318\textwidth]{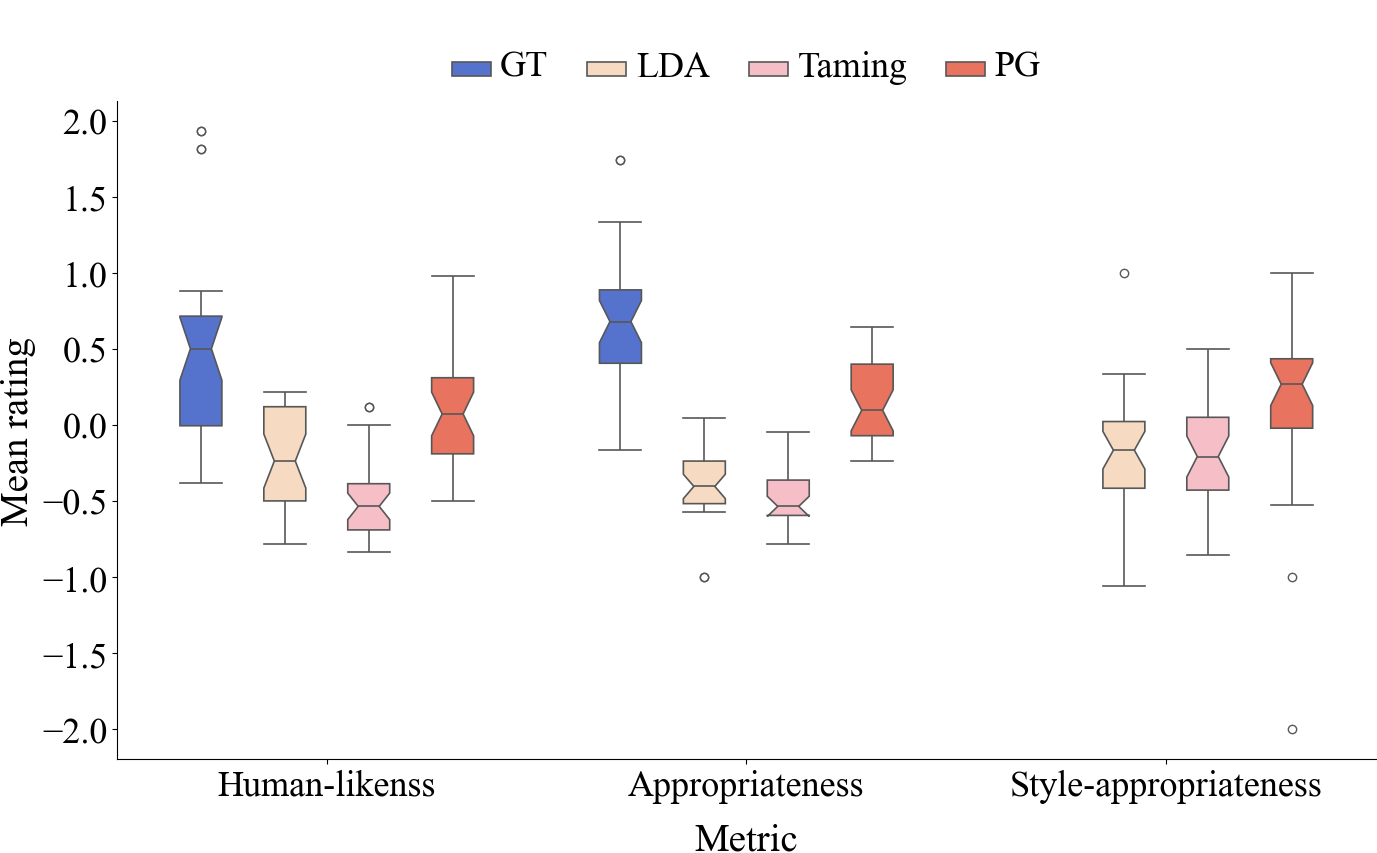}%
    \label{fig:hotmapTriHL} }
    % \hfil
    \subfloat[ZEGGS]{\includegraphics[width=0.333\textwidth]{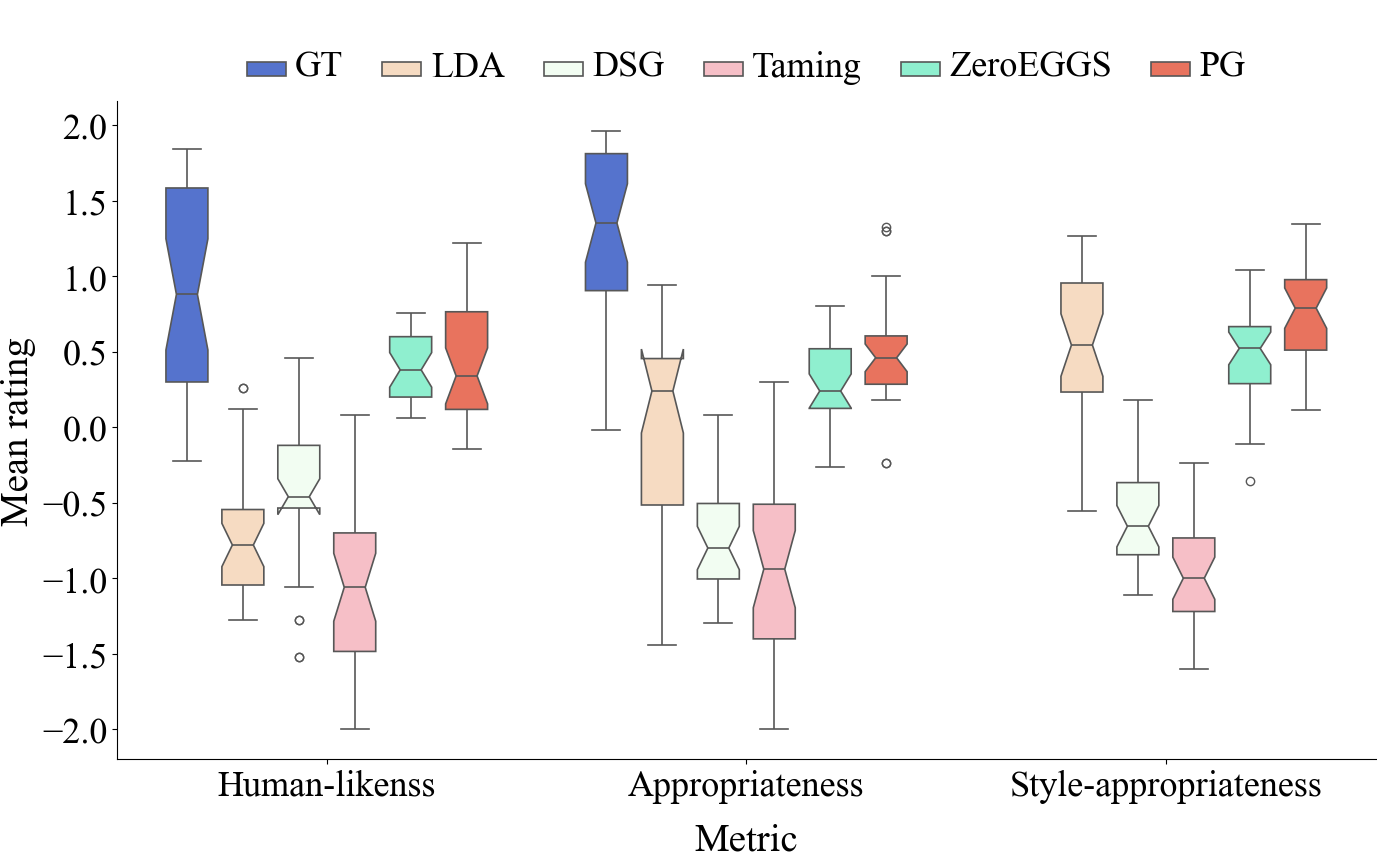}%
    \label{fig:hotmapTriAPP} }
    % \hfil
    \subfloat[BEAT]{\includegraphics[width=0.32\textwidth]{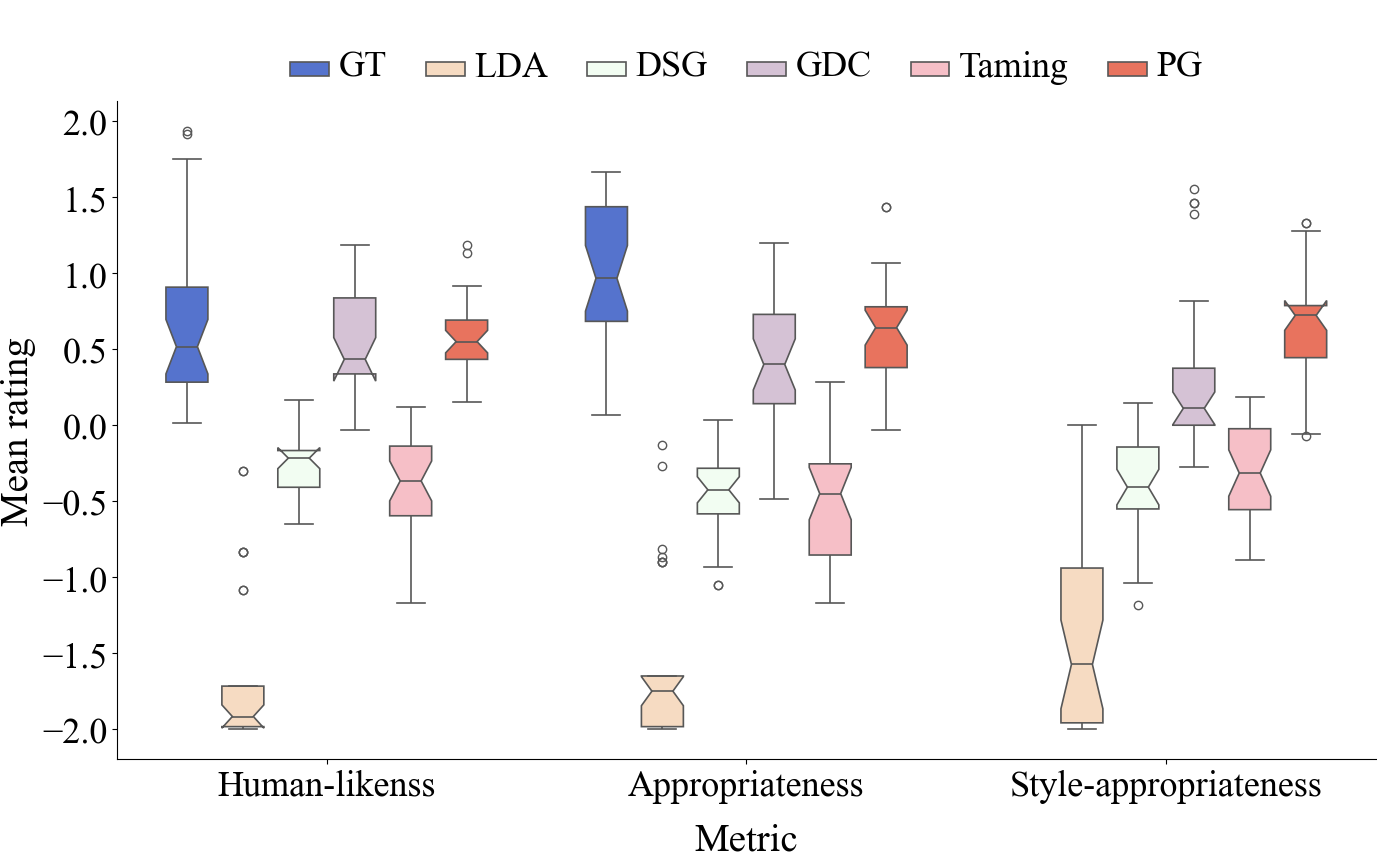}%
    \label{fig:hotmapTriSAPP} }
    \caption{The mean rating of each metric for each approach across the three datasets in comparative experiments.}
\label{fig: compaboxplot}
\end{figure*}

A total of thirty volunteer participants, 17 males and 13 females aged between 19 and 31, were recruited for this study. Among them, 22 participants were Chinese nationals, while the remaining eight were international students from the USA and UK. Notably, all participants in this study exhibited a high level of English proficiency. 

One-way ANOVA and post-hot Tukey multiple comparison tests were conducted to determine if the models' scores differed on the three evaluation aspects. The results are shown in Table \ref{tab:overview} and Figure \ref{fig: compaboxplot}. 
% The post-hoc analysis information is provided in the Appendix \ref{appendix: userStudy}.

The results indicate that the GT achieves the highest scores ($0.51\pm1.73$ and $0.95\pm1.13$) in the Trinity and ZEGGS datasets, exhibiting statistically significant differences ($p<0.001$) in human-likeness evaluations when compared to model-generated gestures. The GT is characterized by a diverse yet limited array of gestures, each with distinct traits that enhance movement realism. However, these gestures belong to the dataset's long-tail distribution, challenging the models' learning capabilities. Additionally, these unique gestures impact the appropriateness and style-appropriateness scores. Conversely, while the GT achieves higher scores ($0.65\pm1.16$), no significant differences were observed compared with the PG ($0.56\pm1.14$) and GDC ($0.54\pm1.12$) in the BEAT dataset analysis. This suggests that these models are more closely aligned with GT benchmarks in this dataset.

The experiments on the Trinity dataset show our proposed model ($0.12\pm1.09$, $0.138\pm1.12$, and $0.203\pm1.06$) outperforming both LDA ($-0.22\pm0.98$, $-0.39\pm1.08$, and $-0.18\pm1.10$) and Taming ($-0.48\pm0.96$, $-0.47\pm1.07$, and $-0.23 \pm 1.05$) architectures significantly ($p < 0.001$) across all metrics. This superior performance is due to the more natural and relaxed gestures produced by our model, PG, enhancing its effectiveness compared to the LDA and Taming models, which fall short in accurately capturing the acoustic rhythm.

% \noindent
\begin{figure*}[!htbp]
    \centering
    \subfloat[Trinity]{\includegraphics[width=0.33\textwidth]{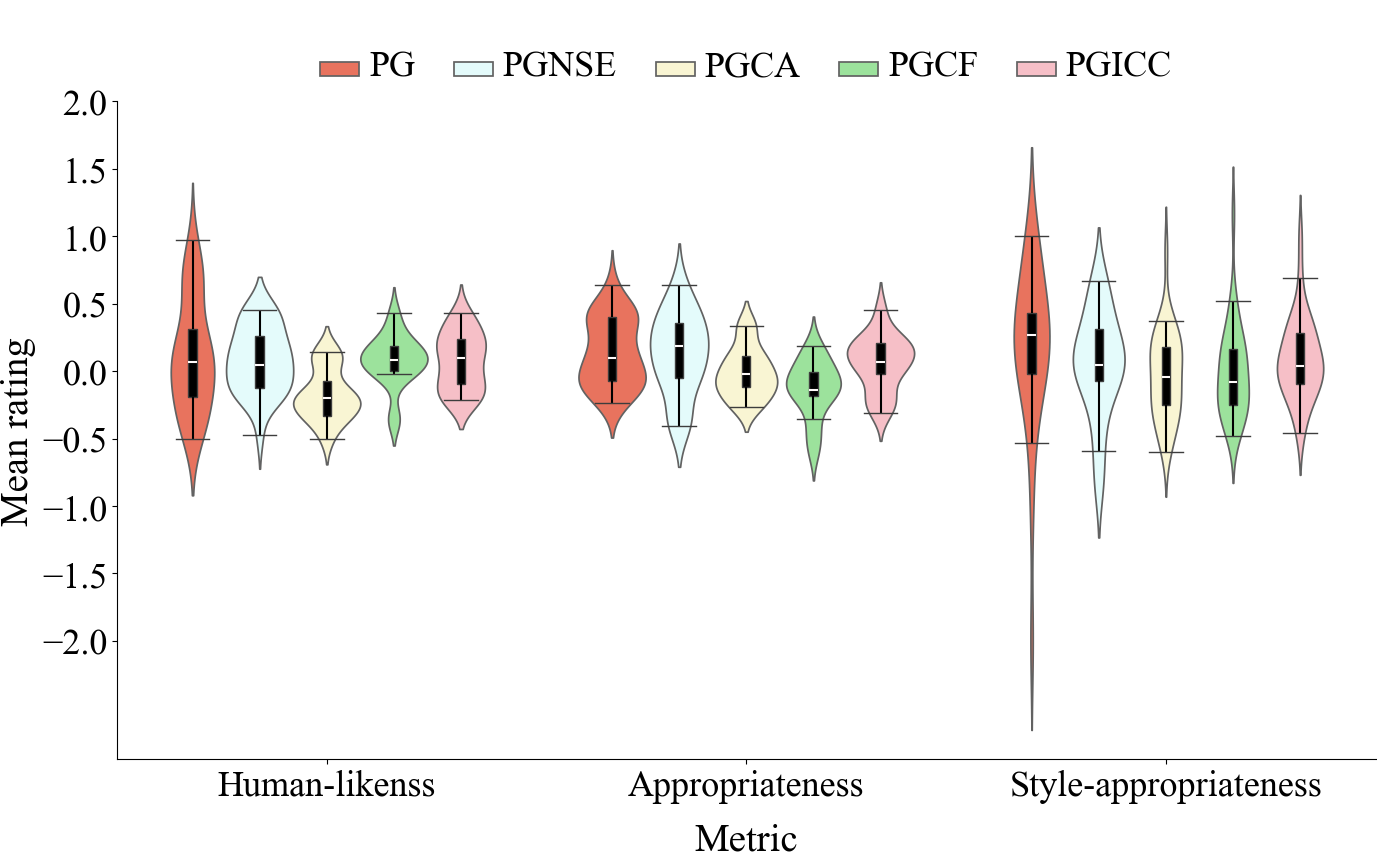}%
    \label{fig:trinity_abla} }
    % \hfil
    \subfloat[ZEGGS]{\includegraphics[width=0.33\textwidth]{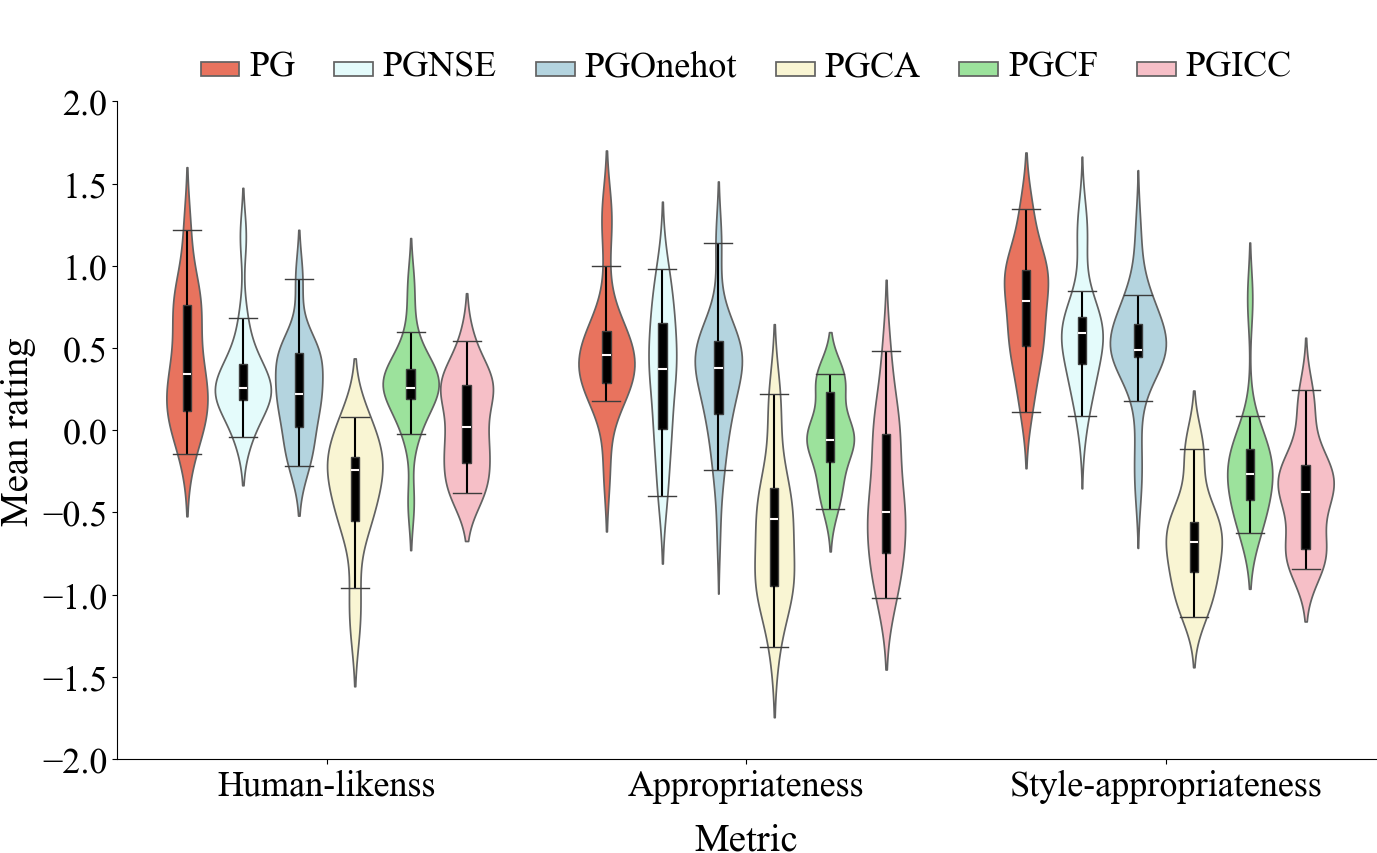}%
    \label{fig:zeggs_abla} }
    % \hfil
    \subfloat[BEAT]{\includegraphics[width=0.33\textwidth]{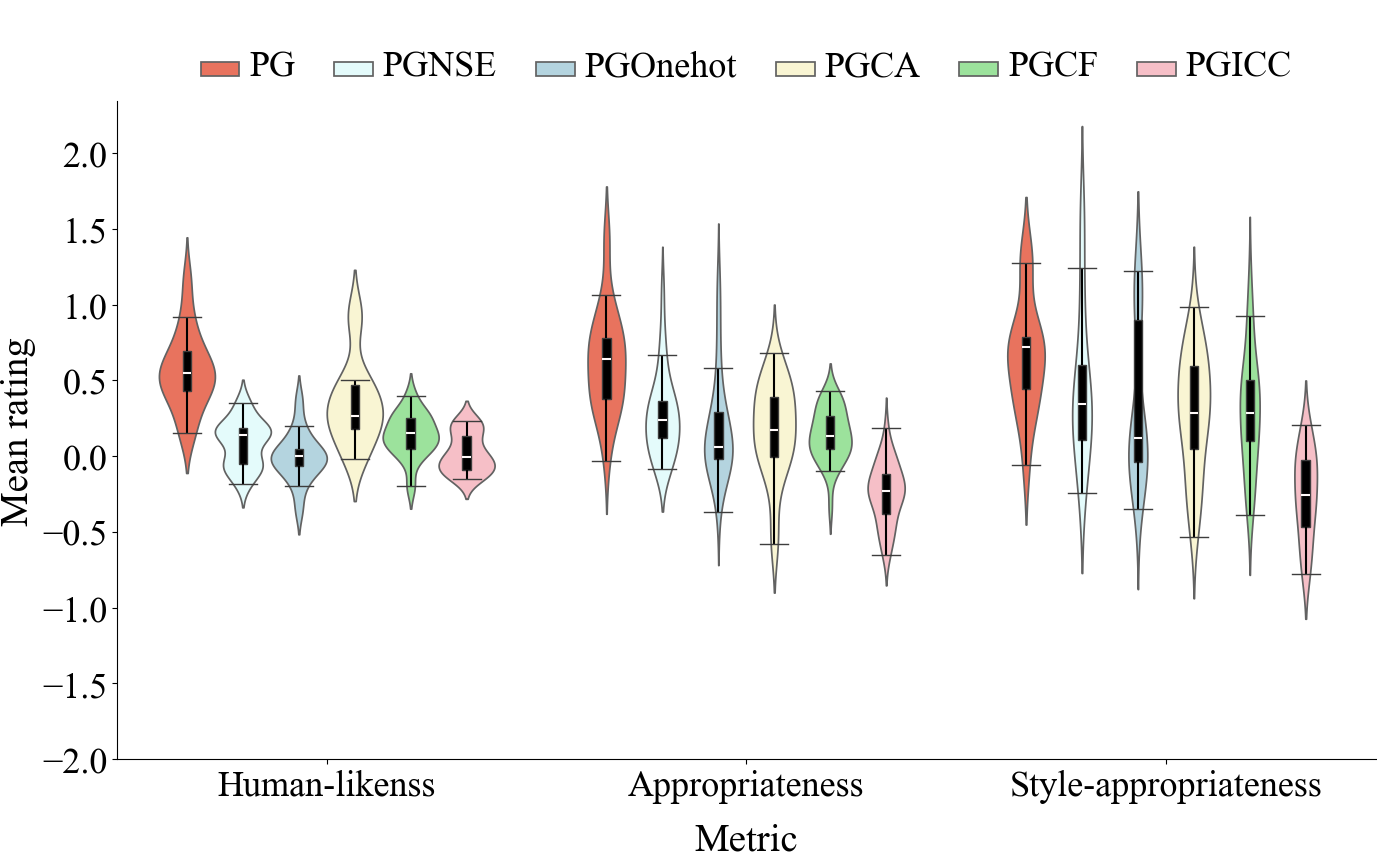}%
    \label{fig:beat_abla} }
    \caption{The mean rating of each metric for each approach across the three datasets in ablation experiments.}
\label{fig: ablaviolin}
\end{figure*}

Evaluation of the ZEGGS dataset showed statistically significant differences ($p < 0.001$) between our method ($0.42\pm1.17$, $0.48\pm1.29$, and $0.76\pm1.34$) and others across all three metrics. However, there was no statistically significant difference ($p>0.05$) between our method and ZeroEGGS ($0.38\pm1.11$) in terms of human likeness, though our method achieved a slightly higher score. These findings suggest that both our proposed model and ZeroEGGS can generate vivid gestures. Our advantage lies in the ability to synthesize emotional gestures solely through audio input, without relying on any reference example animations or labels.

In the BEAT dataset experiments, our PG model exhibited significant improvements ($0.56\pm1.14$, $0.63\pm1.10$, and $0.66\pm1.16$) in three metrics compared to DSG+ ($-0.28\pm1.17$, $-0.49\pm1.15$, and $-0.40\pm1.24$), LDA ($-1.65\pm0.73$, $-1.59\pm0.74$, and  $-1.35\pm1.05$), and Taming ($-0.41\pm1.14$, $-0.52\pm1.14$, and $-0.32\pm1.24$) , reflecting the degradation in synthesis quality observed in DSG+ and LDA when incorporating all styles. While human-likeness metrics were comparable to GDC, the appropriateness metric of PG achieved higher scores than GDC ($0.47\pm1.25$) despite the GDC's ability to better align with speech rhythm. Users reported that gestures generated by GDC overly emphasized prosodic cues, resulting in unnatural and frequent gestures. Furthermore, these gestures often displayed repetitive patterns with limited stylistic diversity, clearly indicating their origin from the GDC model and leading to a lower score ($0.30\pm1.27$) in the style-appropriateness metric. This finding implies that the quality of gestures is not solely determined by accurately matching the audio rhythm.

\subsubsection{Objective Evaluation}\label{sec:objective eval}
We introduce three objective evaluation metrics: Fréchet Gesture Distance (FGD) in both feature and raw data spaces\cite{yoonyoungwoo2020Speecha}, and BeatAlign\cite{li2021ai}. FGD, inspired by the Fréchet Inception Distance (FID)\cite{heusel2017gans}, assesses the quality of generated gestures and demonstrates a moderate correlation with human-likeness ratings, outperforming other objective metrics\cite{kucherenko2023Evaluating}. Additionally, BeatAlign measures gesture-audio synchrony by calculating the Chamfer Distance between audio and gesture beats, providing insights into the temporal alignment of generated gestures with speech rhythms.

Table \ref{tab:overview} displays our results, highlighting the state-of-the-art performance of our method in objective evaluations using FGD and BeatAlign metrics. Our model outperforms (289.42 for Trinity, 28.13 for ZEGGS, and 264.06 for BEAT) other architectures in FGD, effectively generating gestures that align closely with the Ground Truth (GT). It also achieves superior BeatAlign scores (0.69 for Trinity, 0.68 for ZEGGS, and 0.68 for BEAT) compared to other models, except for GDC (0.69 for BEAT), demonstrating its efficacy in producing co-speech gestures that synchronize accurately with speech rhythms. Although GDC scores highest in BeatAlign, corroborating user feedback, its overemphasis on prosodic cues leads to frequent high-frequency gestures. While technically accurate, this diminishes gesture naturalness.

\subsection{Ablation Studies}
Ablation studies were performed to evaluate the impact of key components on our model's efficacy, specifically targeting the global fuzzy feature extractor and Adaptive Layer Normalization (AdaLN). 
\subsubsection{Ablation of Global Fuzzy Feature Extractor}
For the global fuzzy feature extractor, we explored the outcomes of removing this component (we call it: No Style Encoding, PGNSE) and replacing it with One-hot embedding (PGOnehot) for discrete feature extraction. PGOnehot was not applied to the Trinity dataset due to its limited style variability. 

Our analysis of the global fuzzy feature extractor shows no significant differences ($p>0.05$) between PG and PGNSE on the Trinity dataset in three subjective metrics, likely due to its limited range of styles. However, the ZEGGS dataset reveals significant variances in three metrics between PG ($0.42\pm1.17$, $0.48\pm1.29$, and $0.76\pm1.34$) and PGOnehot ($0.25\pm1.19$, $0.33\pm1.31$, and $0.51\pm1.28$), while no notable differences in human-likeness and appropriateness metrics are observed between PG and PGNSE ($0.35\pm1.29$). PG ($0.76\pm1.34$) outperforms PGNSE ($0.59 \pm 1.38$) in style-appropriateness, likely because PGNSE cannot ensure a consistent style throughout the sequence. Conversely, the BEAT dataset exhibits significant differences ($p<0.001$) between PG ($0.56\pm1.14$, $0.63\pm1.10$, and $0.66\pm1.16$) and the other methods, indicating the superior capability of the global fuzzy feature inference mechanism in capturing stylistic nuances. Moreover, while PGOnehot is capable of capturing various logo gesture styles, it may compromise the naturalness of the movements. 

\subsubsection{Ablation of AdaLN}
We integrated Cross-Attention (PGCA), In-Context Conditions (PGICC), and Concatenation of Features (PGCF) into our analysis to evaluate AdaLN's effectiveness. This structured approach enabled a comprehensive assessment of each component's contribution to the model's overall performance and its role in audio-based gesture generation. The implementations of Cross-Attention and In-Context Conditions follow the designs in \cite{peebles2022Scalable}, while Feature Concatenation combines gesture and encoded audio features along the feature axis, a technique proven effective in related studies\cite{zhao_attention-driven_2023}. Separate user studies were conducted for each component, with findings presented in Table \ref{tab:overview} and Figure \ref{fig: ablaviolin}. 

In the ablation studies concerning AdaLN, the replacement of the AdaLN module with alternative architectural frameworks precipitated a significant degradation in performance across all metrics. This reduction in efficacy can be ascribed to the deficiency of alternative architectures in synchronizing speech rhythm and capturing stylistic nuances with precision. This outcome underscores the pivotal role of a uniform mechanism that applies an identical function across all attention layers throughout the sequence. 

\subsection{Generalization and Robustness}
In addition, we test our method's generalization capabilities. We utilized in-the-wild speech audio collected from TED talks. Our system adeptly generates consistent gestures from dataset types and seamlessly produces gestures from untagged, in-the-wild audio. It also showcases remarkable robustness against various auditory disturbances, such as background music, applause, urban noise, and decorative sounds. This adaptability highlights the system's ability to handle a broad spectrum of audio inputs, ensuring the creation of naturalistic gestures despite significant noise interference. Such resilience emphasizes the system's suitability for real-world applications. Yet, we encountered certain inherent challenges in assessing the generalizability and robustness of alternative models during our experimentation. 
% More details can be found in Appendix \ref{app:genandrot}, and supporting videos. 

\section{DISSCUSTION and CONCLUSION}\label{sec:CONCLUSION}
In this work, we introduce \textit{Persona-Gestor}, a novel network architecture designed for the generation of personality gestures, leveraging solely raw speech audio. At its core, \textit{Persona-Gestor} combines a fuzzy feature extractor and an AdaLN transformer diffusion architecture.

The fuzzy feature extractor utilizes a fuzzy feature inference strategy in the dual-component module to implicitly infer both fuzzy stylistic features and specific details embedded within the audio data autonomously. These elements are combined into a unified latent representation, facilitating the generation of speaker-aware personalized 3D full-body gestures. This approach incorporates a highly influential feature into the capability to synthesize personality gestures through automatically inferred fuzzy features, removing the necessity for explicit style labels or additional features. This advancement facilitates the end-to-end generation of gestures that resonate with the speaker's unique characteristics, directly from raw speech audio. Thereby, integrating fuzzy feature inference ensures a seamless and intuitive creation process that enhances generalization and user accessibility.

The AdaLN mechanism is a conditional mechanism that uniformly applies a specific function across all sequence tokens. This strategic incorporation significantly augments the model's proficiency in accurately capturing and representing both conditional dependencies and output characteristics with greater efficiency. We demonstrate that AdaLN also facilitates a refined understanding and processing of the complex interplay between the continuous fuzzy features conditional input and the resultant gesture synthesis, leading to enhanced model performance and output fidelity. Ultimately, \textit{Persona-Gestor} utilizes diffusion mechanism for producing a diverse spectrum of gesture outputs.

Our approach presents multiple benefits: 1) It exclusively uses raw speech audio to synthesize speaker-aware personalized gestures, bypassing the requirement for extra inputs, which enhances user-friendliness. 2) It achieves the full-body (including finger motions and locomotion) gestures' superior synchronization with speech, capturing rhythm, intonation, and certain semantics without compromising naturalness. 3) It showcases improved generalization and robustness, adapting effectively across varied conditions.

Our study highlights key areas for enhancement: Firstly, the model's sole dependence on speech audio may limit its effectiveness in capturing style features within segments of minimal speech. Secondly, the lack of control over the movement path and orientation of the digital human could lead to unintended gestures. Thirdly, our model may not effectively replicate certain gestures, which are crucial for expressing specific states. These observations underscore the necessity for improvements to broaden the model's ability to accurately convey a wide range of human gestures.

\section{ACKNOWLEDGMENTS}
This work was partially supported by the "Pioneer" and "Leading Goose" R\&D Program of Zhejiang (No.2023 C01212), the National Key Research and Development Program of China (No.2022YFF 0902305), the Public Welfare Technology Application Research Project of Zhejiang (No.LGF21F020002, No.LGF22F020008), the Key Program and development projects of Zhejiang Province of China (No.2021C03137), and the Key Lab of Film and TV Media Technology of Zhejiang Province (No.2020E10015).

\bibliography{mybibliography}

\bibliographystyle{IEEEtran}

\vfill
% 附录部分
\clearpage
%\appendix
\appendices
% \appendixpage
% \noappendicestocpagenum
% \addappheadtotoc
% \input{Appendix}
\end{document}